\documentclass[prc,preprint,showpacs,floatfix,nofootinbib,tightenlines]{revtex4}

\usepackage{graphicx}  %
\usepackage{bm}  %
\usepackage{amsmath,amssymb}
\usepackage{braket}

\newcommand{\beqn}{\begin{equation}}
\newcommand{\eeqn}{\end{equation}}
\newcommand{\bea}{\begin{eqnarray}}
\newcommand{\eea}{\end{eqnarray}}
\newcommand{\ba}{\begin{align}}
\newcommand{\ea}{\end{align}}

\newcommand{\ad}{a^\dagger}
\newcommand{\flow}{s}
\newcommand{\Hzero}{T}

\newcommand{\vlowk}{V_{{\rm low}\,k}}
\def\vaa{\langle a | V_{2 \flow} | a \rangle}
\def\vbb{\langle b | V_{2 \flow} | b \rangle}
\def\vab{\langle a | V_{2 \flow} | b \rangle}
\def\vba{\langle b | V_{2 \flow} | a \rangle}
\def\vcc{\langle c | V_{2 \flow} | c \rangle}
\def\waa{\langle \alpha | V_{3 \flow} | \alpha \rangle}
\def\wbb{\langle \beta | V_{3 \flow} | \beta \rangle}
\def\wab{\langle \alpha | V_{3 \flow} | \beta \rangle}
\def\wba{\langle \beta | V_{3 \flow} | \alpha \rangle}

\newcommand{\mybar}[1]{\overline{{#1}}}

\begin{document}


\title{Three-Body Forces Produced by a Similarity Renormalization 
          Group Transformation in a Simple Model}

\author{S.K.\ Bogner}\email{bogner@nscl.msu.edu}

\affiliation{National Superconducting Cyclotron Laboratory 
       and Department of Physics and Astronomy,
       Michigan State University, East Lansing, MI 48824}

\author{R.J.\ Furnstahl}\email{furnstahl.1@osu.edu}
\author{R.J.\ Perry}\email{perry.6@osu.edu}

\affiliation{Department of Physics,
         The Ohio State University, Columbus, OH 43210}

\date{August, 2007}

\begin{abstract}
A simple class of unitary renormalization group transformations that force
hamiltonians towards a band-diagonal form produce few-body
interactions in which low- and high-energy states are decoupled, which 
can greatly simplify many-body calculations. 
One such
transformation has been applied to phenomenological and effective field
theory nucleon-nucleon interactions with success, but further progress
requires consistent treatment of at least
the three-nucleon interaction. 
In this paper we demonstrate in an extremely simple model
how these renormalization group
transformations consistently evolve two- and three-body interactions
towards band-diagonal form, and introduce a diagrammatic approach that
generalizes to the realistic nuclear problem.
\end{abstract}

\smallskip

\maketitle


\section{Introduction}

Wilsonian renormalization group transformations are designed to replace
explicit coupling between disparate distance or energy scales with
effective interactions in which disparate scales are decoupled \cite{wilson:1975}.  
In recent work we have employed a simple
unitary renormalization group transformation to study the
nucleon-nucleon (NN) interaction~\cite{Bogner:2006srg,Bogner:2007jb,srgwebsite}. This
transformation is a simplified version of Wegner's flow
equations~\cite{Wegner:1994} and one of a much larger class of
Similarity Renormalization Group (SRG) transformations developed by
Glazek and Wilson~\cite{Glazek:1993rc}.
The transformation leads to NN potentials for which calculations
of few-nucleon binding
energies and other observables converge 
rapidly~\cite{Bogner:2006srg,Bogner:2007jb,NCSMSRG} .
However, further progress will require the consistent treatment
of at least the three-nucleon interaction.

The basics of the unitary evolution
are simply stated.
Consider a unitary transformation $U_\flow$ of a hamiltonian $H$, 
\beqn
   H_\flow = U_\flow H U^\dagger_\flow ,
\eeqn
where $\flow$ is a continuous flow parameter. We take $s=0$ for the
initial value with $U_0$ the identity transformation, 
so $H$ is our input hamiltonian. We want to choose
$U_\flow$ so that $H_\flow$ is diagonalized (band-diagonalized in more
realistic cases)  within a specified basis as $\flow \rightarrow
\infty$, which will realize the desired decoupling of low- and high-energy states.
Direct differentiation shows that $H_\flow$ evolves according to
\beqn
  \frac{dH_\flow}{d\flow}
    = [\eta(\flow),H_\flow] \; ,
\eeqn
with
\beqn
   \eta(\flow) = \frac{dU_\flow}{d\flow} U^\dagger_\flow 
          = -\eta^\dagger(\flow)
   \; .
\eeqn
Choosing $\eta(\flow)$ specifies the unitary transformation.
A simple choice is
\beqn
  \eta(\flow) =  [T, H_\flow]
  \label{eq:choice}
   \; ,
\eeqn
where $T$ is a fixed matrix (independent of $s$), which gives the flow
equation,
\beqn
  \frac{dH_\flow}{d\flow} 
  =  [ [T, H_\flow], H_\flow] \; .
  \label{eq:commutator}
\eeqn 
In Refs.~\cite{Bogner:2006srg,Bogner:2007jb,Szpigel:2000xj} $T$ was chosen to be
the kinetic energy; here we choose the hamiltonian for free particles in an
infinite square well. For explicit calculations we employ a basis in which $T$
is diagonal and it is in this representation that $H_\flow$ is driven towards
band-diagonal form, as we will see below. 

A principal advantage of SRG transformations is that all operators are
consistently transformed, which means that all observables are
invariant. For the simple unitary transformation this is obvious, simply because it is
unitary. An additional advantage is that SRG transformations
readily handle Fock space operators; indeed, Glazek and Wilson designed
them to attack light-front quantum chromodynamics. Interactions that
change particle number are not required for low-energy nuclear physics, but we 
do require the consistent evolution of all
many-body operators. 
If we express the hamiltonian in terms of creation and destruction
operators, it is evident that the commutators in
Eq.~(\ref{eq:commutator}) will generate many-body interactions even
if $H$ includes only a two-body interaction.
In principle this could make practical calculations intractable, but 
in applications of interest (e.g., to nuclear physics) we
can choose transformations that maintain a hierarchy of many-body forces
such that for sufficiently dilute many-body systems we only need to
evolve few-body operators.

In this paper we illustrate how a unitary SRG consistently
evolves two-body \emph{and} three-body interactions. We choose what may be the
simplest possible example of such evolution, bosons in a
two-level system in which only two states are coupled in each of the
two-particle and three-particle sectors.  
As a specific realization, we use a
one-dimensional infinite square well with the Fock space truncated to
allow only the two lowest modes. 
Despite the simplicity of the model, it exhibits many of the
basic issues to be confronted in
evolving few-body interactions
and illustrates a diagrammatic formulation of the SRG equations
that carries over
to more realistic problems (such as nuclei).

In the severely truncated model space, the largest sub-matrix we will
encounter is $2\times 2$; so we start in Sect.~\ref{sec:2by2} by
reviewing the unitary transformation for the simple case of $2\times 2$
matrices. 
In Sect.~\ref{sec:bosons}
we explicitly show in our simple example that the
unitary evolution of hamiltonians containing two-body interactions must
produce three-body interactions.
We derive the explicit
$2\times 2$ matrices that represent the flowing two-body interaction,
$V_{2s}$, and the positive parity part of the flowing three-body
interaction, $V_{3s}$.  In Sect.~\ref{sec:diagrams} we develop simple
diagrammatic rules for the production of evolution equations governing
any $N$-body interaction, $V_N$, and show that these readily reproduce
the equations for $V_2$ and $V_3$ in our simple model. The diagrammatic
representation suggests several approximations that are tested.
We summarize and make connections to nuclear physics and other
problems of interest in Sect.~\ref{sec:summary}.


\section{$2\times 2$ Matrix Evolution}
\label{sec:2by2}

Let us start by considering a two-state system,
with the hamiltonian 
represented by a $2\times 2$ hermitian matrix. 
In realistic many-body quantum mechanics problems, we frequently
encounter a hamiltonian that naturally divides, $H = T + V$, with $T$ a
one-body operator.%
\footnote{There is considerable freedom in the choice of $T$, it need
not even be a one-body operator. $V$ can be initially chosen to be a
two-body operator, but the commutators will then automatically generate
three-body operators, four-body operators, etc.}
In this case, the unitary transformation
\beqn
   H_\flow = U_\flow \Hzero U^\dagger_\flow \equiv T + V_\flow 
\eeqn
\emph{defines} the evolved interaction $V_\flow$. 
In terms of basis states $\ket{i}$, where $T \ket{i} = E_i \ket{i}$, the flow
equation is
\beqn
\frac{d}{ds} V_{ij} = -(E_i-E_j)^2 V_{ij} + \sum_k (E_i+E_j-2 E_k) V_{ik} V_{kj} \;.
\eeqn
Here the renormalization flow parameter, $s$, has the dimension of
inverse energy squared. The first term on the right-hand-side drives
$V$, and therefore $H$, towards band-diagonal form as $s$ increases
from zero.

In our two-state example we will also take $H=\Hzero+V$
and use eigenstates of $\Hzero$ as a basis, with $\Hzero | 1 \rangle =
E_1 | 1 \rangle$ and $\Hzero | 2 \rangle = E_2 | 2 \rangle$.  In this
representation,
\beqn
\Hzero = \begin{pmatrix}  E_1 & 0 \\ 0 & E_2 \end{pmatrix} \;,
\eeqn
and the full hamiltonian is
\beqn
H_s = \begin{pmatrix}  E_1+V_{11}(s) & V_{12}(s) \\ 
    V_{21}(s) & E_2+V_{22}(s) \end{pmatrix} \;,
\eeqn
where $V_{ij}(s) \equiv \langle i | V_s | j \rangle$, which we take
to be real.
Any real symmetric $2\times 2$ matrix can be expressed using the
identity matrix, $I$,
and the Pauli spin matrices, $\sigma_x$ and $\sigma_z$,
\beqn
T = {\cal E} I + \Omega \sigma_z,
\eeqn
\beqn
V_\flow = c I + \omega_z(s) \sigma_z + \omega_x(s) \sigma_x  \;,
\eeqn
where ${\cal E}=\frac{1}{2} (E_1+E_2)$, $\Omega=\frac{1}{2}(E_1-E_2)$, 
$c=\frac{1}{2}(V_{11}+V_{22})$, $\omega_z(s)=\frac{1}{2}(V_{11}-V_{22})$ 
and $\omega_x(s)=V_{12}=V_{21}$. Since $I$ commutes with all other matrices, 
${\cal E}$ and $c$ drop out of the flow equations and $c$ is therefore independent of $s$. 
So, we need only consider the Pauli
spin matrices $\sigma_z$ and $\sigma_x$ to understand the real $2\times 2$
hamiltonian. If an additional angle is introduced using $\sigma_y$, it turns out to be
another flow constant that produces no qualitative change in the way the
hamiltonian flows.

Dropping the constants ${\cal E}$ and $c$, which can simply be added
back in after the
flow equations are solved, we work with
\beqn
  V_\flow = \omega_z(s) \sigma_z + \omega_x(s) \sigma_x  \;,
\eeqn
and
\beqn
  T = \Omega \sigma_z  .
\eeqn
With the above matrices, the unitary renormalization group flow equation 
(\ref{eq:commutator}) becomes
\beqn
  \frac{d}{ds}\bigl[\omega_z(s) \sigma_z + \omega_x(s) \sigma_x\bigr] =
  4 \Omega\bigl[\omega_x^2(s) \sigma_z - (\Omega+\omega_z(s)) \omega_x(s) \sigma_x\bigr]
  \;,
\eeqn
or, projecting out coefficients of the Pauli matrices,
\beqn
  \frac{d}{ds}\omega_z(s)= 4 \Omega \omega_x^2(s) \;,  \qquad
  \frac{d}{ds}\omega_x(s)= -4 \Omega (\Omega+\omega_z(s)) \omega_x(s) \;.
\eeqn
These equations illustrate a tremendous advantage of such unitary
transformations in that the exact renormalization group equations are only
second order in $V$. Note that if $|\Omega| >> |\omega_z(s)|$, $\omega_x(s)$ is
obviously driven to zero exponentially.

The equations are easily solved analytically. We introduce the angle $\theta$
and set $\Omega+\omega_z = \omega\cos\theta$ and $\omega_x=\omega\sin\theta$, 
where $\omega=  \sqrt{(\Omega+\omega_z)^2 + \omega_x^2}$. With this change of
variables we find that $\omega$ is a flow constant, so only the angle $\theta$
depends on $\flow$. The flow equation becomes:
\beqn
\frac{d \theta}{ds} = -4 \Omega \omega \sin\theta 
  \;.
\eeqn
Integrating this equation we obtain:
\beqn
 \tan\left(\frac{\theta(s)}{2}\right)
  = \tan\left(\frac{\theta(0)}{2}\right) \, e^{-4 \Omega \omega s} \;.
\eeqn
Thus $\theta$ is driven exponentially either to zero or to $\pi$, depending on
the sign of $\Omega$. In either case the matrix is driven to diagonal form,
with the states in the same order as in $\Omega \sigma_z$.


\section{Bosons in a Square Well Truncated to Two Modes}
\label{sec:bosons}

In this section we consider a simple concrete example
that lets us examine SRG evolution in two-boson and three-boson sectors of Fock
space.
We choose $T$ to be the hamiltonian for
non-interacting bosons in an infinite square well where $-L/2 \le x \le
L/2$,
\beqn
T=\frac{P^2}{2m}+V_{\rm well} \;.
\eeqn 
We work in the $T$ eigenstate basis, 
\beqn
 \phi_n(x)=\sqrt{\frac{2}{L}}~{\rm sin}\left(\frac{n \pi (x-L/2)}{L}\right) \;,
\eeqn
and truncate the boson Fock space to allow only the
ground and first excited state modes. We will see that this drastic
truncation reduces the two- and three-boson problems to
$2\times 2$ matrix problems, as solved in Sect.~\ref{sec:2by2}.

After truncating to two modes, the complete set of eigenvalues of
$\Hzero$ are
\beqn
 E_1 = \frac{\pi^2}{2 m L^2} \;, \qquad E_2= 4 E_1 \;.
\eeqn
 We can build many-body interactions using Fock-space operators and
below we develop a Fock space diagrammatic analysis exploiting the
simple algebra of these operators, but for pedagogical reasons
we first compute matrix elements
of the hamiltonian with two-body and three-body interactions
mechanically. The interactions are given by their matrix elements
between unsymmetrized two- and three-boson states, which must be related
to matrix elements between properly symmetrized states. We choose an
interaction that preserves parity and need only consider the
three-dimensional two-boson space and the two-dimensional positive
parity three-boson space to display the method.
 
It is convenient to use several bases for the many-boson problem,
which we distinguish by the type of bracket used. Unsymmetrized three-boson
states, for example, are $| n_1 n_2 n_3)$, in which $n_i$'s denote the
individual particle square-well states. We also use symmetrized states
$\ket{N_1 N_2}$, in which $N_1$ is the number of bosons in the $n=1$
state and $N_2$ the number in the $n=2$ state. The only two-boson
symmetrized  states are $\ket{20}\equiv\ket{a}$, $\ket{02}\equiv\ket{b}$, and
$\ket{11}\equiv\ket{c}$ and we only need the positive parity symmetric
three-boson states, $\ket{30}\equiv\ket{\alpha}$ and
$\ket{12}\equiv\ket{\beta}$.  

We display the two-boson and three-boson states we need
in Tables \ref{tab:1}--\ref{tab:4}, listing their
free energy and parity.  We also
list the symmetric states' representations in terms of unsymmetrized
states, which we need to compute the initial hamiltonian. If the
initial two-body interaction conserves parity, parity will be conserved
in the evolution, so even and odd parity states will not mix. In general, any
operator that commutes with $T$ and $H$ will commute with $H_\flow$ and result
in an explicit symmetry. Parity conservation
simplifies both the two- and three-boson problems, reducing them to
$2\times 2$ matrix problems. 

\begin{table}[p] 
 \caption{\bf Two-boson unsymmetrized states}
  \renewcommand{\tabcolsep}{10pt}
 \centering 
 \begin{tabular}{|c|c|c|c|} \hline
$|  n_1 n_2)$ & $E_{n_1,n_2}$ & parity \\ \hline \hline
$| 11)$ & $2 E_1$ & $+$ \\ \hline
$| 22)$ & $8 E_1$ & $+$ \\ \hline
$| 12)$ & $5 E_1$ & $-$ \\ \hline 
$| 21)$ & $5 E_1$ & $-$ \\ \hline
\end{tabular}
\label{tab:1}
\vspace*{.4in}
 \caption{\bf Two-boson symmetrized states}
  \renewcommand{\tabcolsep}{10pt}
 \centering 
 \begin{tabular}{|c|c|c|c|c|c|} \hline
 $| N_1 N_2\rangle$ & label & $| n_1 n_2)$-basis&$E_{N_1,N_2}$ & parity \\ \hline \hline
$| 20\rangle$ &$| a \rangle$ &$| 11)$  &$2 E_1$ & $+$ \\ \hline
$| 02\rangle$ &$| b \rangle$ &$| 22)$  & $8 E_1$ & $+$ \\ \hline
$| 11\rangle$ &$| c \rangle$ &$\frac{1}{\sqrt{2}}\bigl[| 12)+ | 21)\bigr]$ 
          & $5 E_1$ & $-$ \\ \hline 
\end{tabular}
\label{tab:2}
\vspace*{.4in}
 \caption{\bf Positive parity three-boson unsymmetrized states}
  \renewcommand{\tabcolsep}{10pt}
 \centering 
 \begin{tabular}{|c|c|c|cl} \hline
$| n_1$ $n_2$ $n_3)$ & $E_{n_1,n_2,n_3}$ & parity \\ \hline \hline

$| 111)$ & $3 E_1$ & $+$ \\ \hline 
$| 122)$ & $9 E_1$ & $+$ \\ \hline
$| 212)$ & $9 E_1$ & $+$ \\ \hline
$| 221)$ &   $9 E_1$ & $+$ \\ \hline
\end{tabular}
\label{tab:3}
\vspace*{.4in}
 \caption{\bf Three-boson symmetrized states}
  \renewcommand{\tabcolsep}{10pt}
 \centering 
 \begin{tabular}{|c|c|c|c|c|cl} \hline
 $| N_1 N_2\rangle$ & label & $| n_1 n_2 n_3)$-basis& $E_{N_1,N_2}$ & parity \\ \hline \hline
$| 30\rangle$ & $| \alpha \rangle $ &$| 111)$&  $3 E_1$ & $+$ \\ \hline
$| 12\rangle$ & $| \beta \rangle $ &$\frac{1}{\sqrt{3}} \bigl[|
122)+| 212) +| 221) \bigr]$& $9 E_1$ & $+$ \\ \hline
$| 21\rangle$ & $| \gamma \rangle $ &$\frac{1}{\sqrt{3}}
\bigl[| 112)+| 121) +| 211) \bigr]$& $6 E_1$ & $-$ \\ \hline
$| 03\rangle$& $| \delta \rangle $&$| 222)$& $12 E_1$ & $-$ \\ \hline
\end{tabular}
\label{tab:4}
\vspace*{.4in}
\caption{\bf Initial matrix elements of $V_2$}
  \renewcommand{\tabcolsep}{10pt}
 \centering 
 \begin{tabular}{|c|c|c|c|c|c|} \hline
$\braket{a|V_{2}|a}=\braket{b|V_{2}|b}=3 g$ \\ \hline
$\braket{a|V_{2}|b}=2 g$ \\ \hline
$\braket{c|V_{2}|c}=4 g$ \\ \hline
$\braket{\alpha|V_{2}|\alpha}=9 g$ \\ \hline
$\braket{\beta|V_{2}|\beta}=11 g$ \\ \hline
$\braket{\alpha|V_{2}|\beta}=2 \sqrt{3} g$ \\ \hline
\end{tabular}
\label{tab:5}
\end{table}

We will see that we must allow for an explicit three-body interaction, $V_3$, without
which the transformation cannot be unitary. 
If we set $V_3=0$ at $s=0$, we will see
that the unitary transformation generates a three-body interaction that maintains
unitarity in the three-boson sector of Fock space. That
is, we need $V_\flow = V_{2 \flow}+V_{3 \flow}$ here. In realistic
cases we need to choose the transformation so that the induced part
of these many-body
interactions remains short-ranged, which for fermions should allow us to
truncate the tower of many-body operators and accurately solve
many-fermion problems with evolved few-body operators only. The
calculation here is easily extended to the four-boson sector of Fock
space, where $V_{4 \flow}$ first appears, and so on. Restricting our
attention to the two-boson and three-boson sectors, we must solve
\beqn
  \frac{dH_\flow}{d\flow} 
  =  [ [\Hzero, V_{2 \flow}+V_{3 \flow}], \Hzero + V_{2 \flow}+V_{3 \flow}] \ .
  \label{eq:commutatorboth}
\eeqn
Solving this in the two-boson sector completely determines $V_{2 \flow}$
and the three-boson sector then determines $V_{3 \flow}$. Since particle
number is conserved, we simply need to take two-boson or three-boson
matrix elements of the flow equation and insert a complete set of
states on the right-hand-side to obtain coupled nonlinear equations for
each independent matrix element.

Two-boson matrix elements of the flow equation determine $V_{2 \flow}$,
\bea
\frac{d}{ds} \vaa &=& -12 E_1 \vab \vba \;,
 \label{eq:vaa} \\
\frac{d}{ds} \vbb &=& 12 E_1\vba \vab \;,
 \label{eq:vbb} \\
\frac{d}{ds} \vab &=& -36 E_1^2 \vab 
 +6 E_1 \bigl[ \vaa - \vbb \bigr]  \vab \;.
 \label{eq:vab}
\eea
There is no coupling between the positive and negative parity states for the interaction
we will study and
$ \vcc$ does not evolve because there is only one odd-parity two-boson
state. We have seen in Sect.~\ref{sec:2by2} 
that these equations have an analytic
solution;  to apply it we need only remove the
part of $H$ that is proportional to an identity matrix and determine $\Omega$,
$\omega_z(0)$ and $\omega_x(0)$ from what remains.

The three-boson evolution equations are readily computed using 
eq.~(\ref{eq:commutatorboth}),
\bea
\frac{d}{ds} \langle \alpha | V_{2 \flow}+V_{3 \flow} | \alpha \rangle 
  &=& 
  -12 E_1\langle \alpha | V_{2 \flow}+V_{3 \flow} | \beta \rangle  \langle \beta 
    | V_{2 \flow}+V_{3 \flow} | \alpha
\rangle  \;,
 \label{eq:valpha} \\
\frac{d}{ds} \langle \beta | V_{2 \flow}+V_{3 \flow} | \beta \rangle 
   &=& 
   12 E_1\langle \beta | V_{2 \flow}+V_{3 \flow} | \alpha \rangle  \langle \alpha 
     | V_{2 \flow}+V_{3 \flow} | \beta
\rangle  \;,
 \label{eq:vbeta} \\
\frac{d}{ds} \langle \alpha | V_{2 \flow}+V_{3 \flow} | \beta \rangle     
  &=& -36 E_1^2  \langle \alpha | V_{2 \flow}+V_{3 \flow} | \beta \rangle \nonumber \\ 
&& \null + 6 E_1  \langle \alpha | V_{2 \flow}+V_{3 \flow} 
| \alpha \rangle  \langle \alpha | V_{2 \flow} +V_{3 \flow} | \beta \rangle \nonumber
 \\ && \null -6 E_1  \langle \alpha | V_{2 \flow}+V_{3 \flow} | \beta \rangle  \langle 
 \beta | V_{2 \flow}+V_{3 \flow} | \beta \rangle
   \;.
  \label{eq:valphabeta}
\eea
Here again we have a $2\times 2$ matrix problem, so it also can be solved
as shown in Sec.~\ref{sec:2by2}. 
One way to find $V_3$ is to solve for $V_2+V_3$ and then
subtract matrix elements of $V_2$ that result from solving for
$V_2$ in the two-boson sector. Instead, we will
discuss a seemingly more complicated procedure that is required in more
realistic examples where matrix elements of $dV_2/ds$ in the three-boson
sector produce momentum delta-functions due to non-interacting spectators. 

We need symmetrized three-boson matrix elements of $V_{2 \flow}$ in
terms of its symmetrized two-boson matrix elements, in order to embed
the solution of eqs.~(\ref{eq:vaa})--(\ref{eq:vab})  in
eqs.~(\ref{eq:valpha})--(\ref{eq:valphabeta}). These are 
computed using unsymmetrized states given in Table~\ref{tab:3}:
\bea
  \langle \beta | V_{2 \flow} | \beta \rangle &=& \vbb +2 \,\vcc \;,
  \\
  \langle \beta | V_{2 \flow} | \alpha \rangle &=& \sqrt{3} \, \vba \;,
  \\
  \langle \alpha | V_{2 \flow} | \alpha \rangle &=& 3 \,\vaa \;.
\eea
We can now rewrite all of the three-boson matrix elements of $V_{2
\flow}$ in terms of its two-boson matrix elements, then use the solution
to eqs.~(\ref{eq:vaa})--(\ref{eq:vab}) to rewrite the three-boson equations for $V_{3
\flow}$:
\bea
\frac{d}{ds} \waa &=& -12 E_1 \bigl[ \sqrt{3} \vab  \wba+ \sqrt{3}           
  \wab  \vba \nonumber \\
&&  \qquad\qquad \null + \wab  \wba \bigr] \;,
  \label{eq:waa}  
\eea
\bea
\frac{d}{ds} \wbb &=& 12 E_1 \bigl[ 2 \vba \vab \nonumber \\
  && \qquad\qquad \null + \sqrt{3} \vba  \wab+ \sqrt{3}   \wba  \vab        
  \nonumber \\
  && \qquad\qquad \null+ \wba  \wab \bigr]   \;,
  \label{eq:wbb}  
\eea
\bea
  \frac{d}{ds} \wab &=& -36 E_1^2  \wab 
   + 12 E_1 \sqrt{3} ( \vaa-\vcc) \vab \nonumber \\
  &&
    +6 E_1 \bigl[ (\waa-\wbb) \sqrt{3} \vab+ \nonumber \\
    && \qquad\qquad\null(3 \vaa-\vbb-2 \vcc) \wab \bigr] \nonumber \\
  && \qquad\null
     - 6 E_1 (\waa-\wbb ) \wab  \;.
  \label{eq:wab}
\eea
There is no obvious simplification here, quite the contrary. Replacing matrix elements of
$V_2$ with their solutions from the two-boson sector complicates the equation, but we
will
return to this issue after developing a diagrammatic analysis. It
is no longer obvious that this is another $2\times 2$ matrix example
and it is far more tedious to find an analytic solution, but once again
the entire hamiltonian is driven to diagonal form, which requires $V_2$
and $V_3$ to be separately driven to diagonal form.

\begin{figure*}[thb]
  \includegraphics*[width=3.05in]{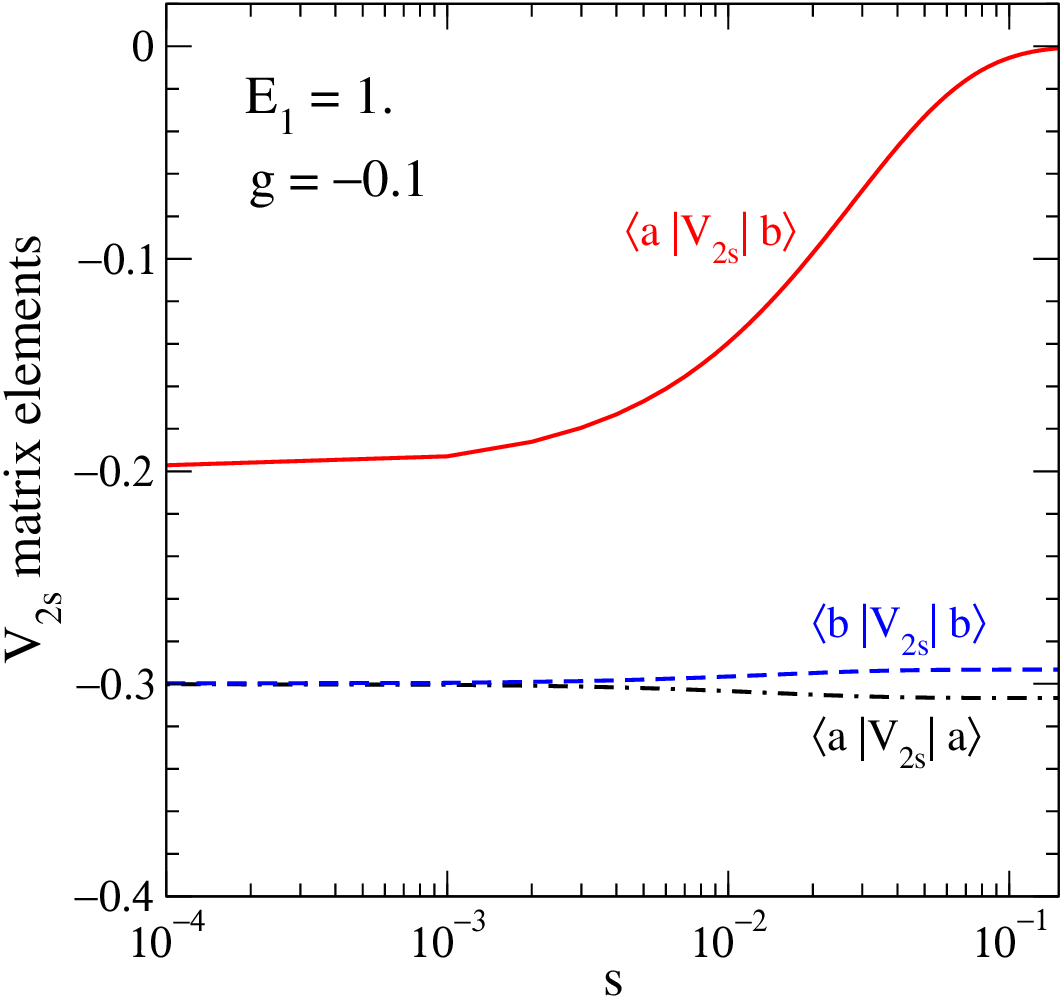}
  \hspace*{.1in}
  \includegraphics*[width=3.1in]{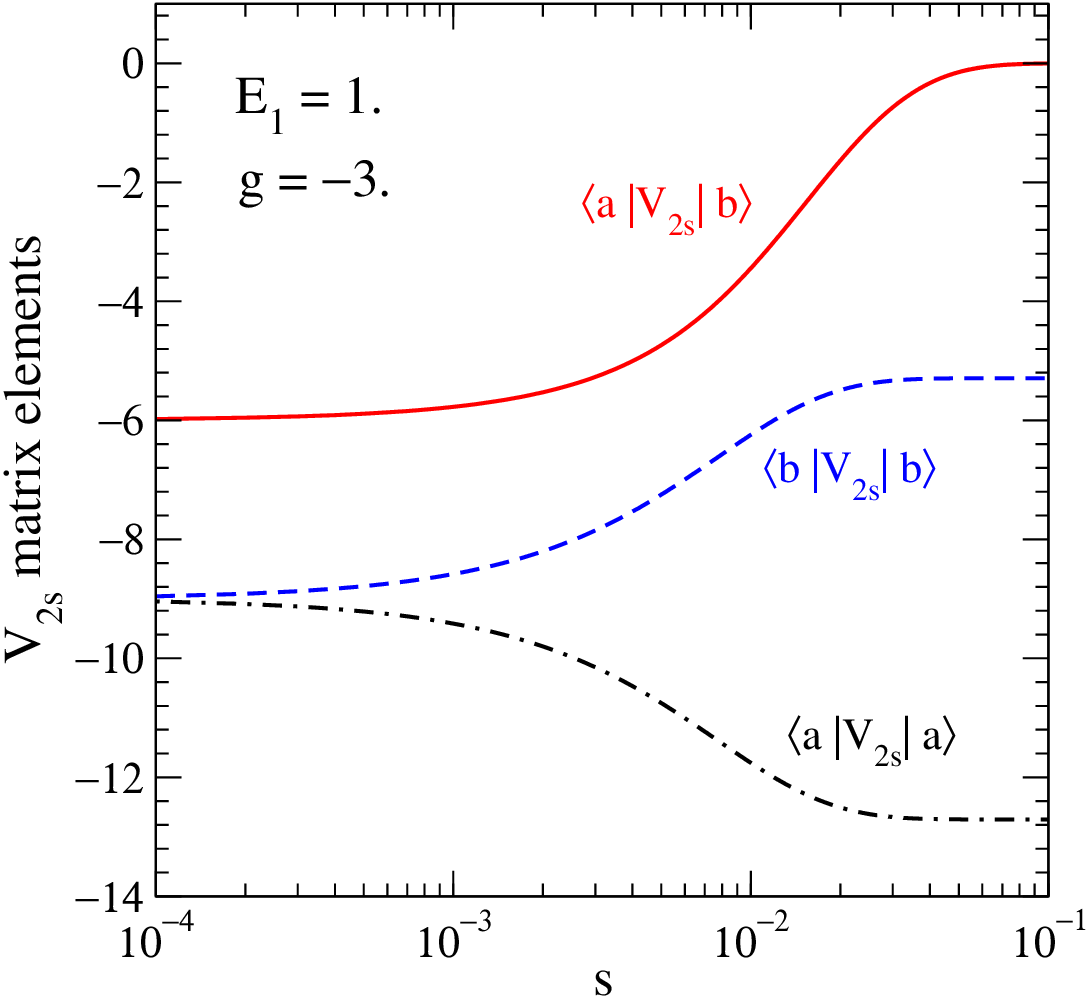}
  \caption{Hamiltonian matrix elements of $V_2$ as a function of the flow
  parameter $s$ in the two-boson sector for weak and strong couplings.}
  \label{fig:v2matrix}
  \vspace*{.1in}
  \includegraphics*[width=3.05in]{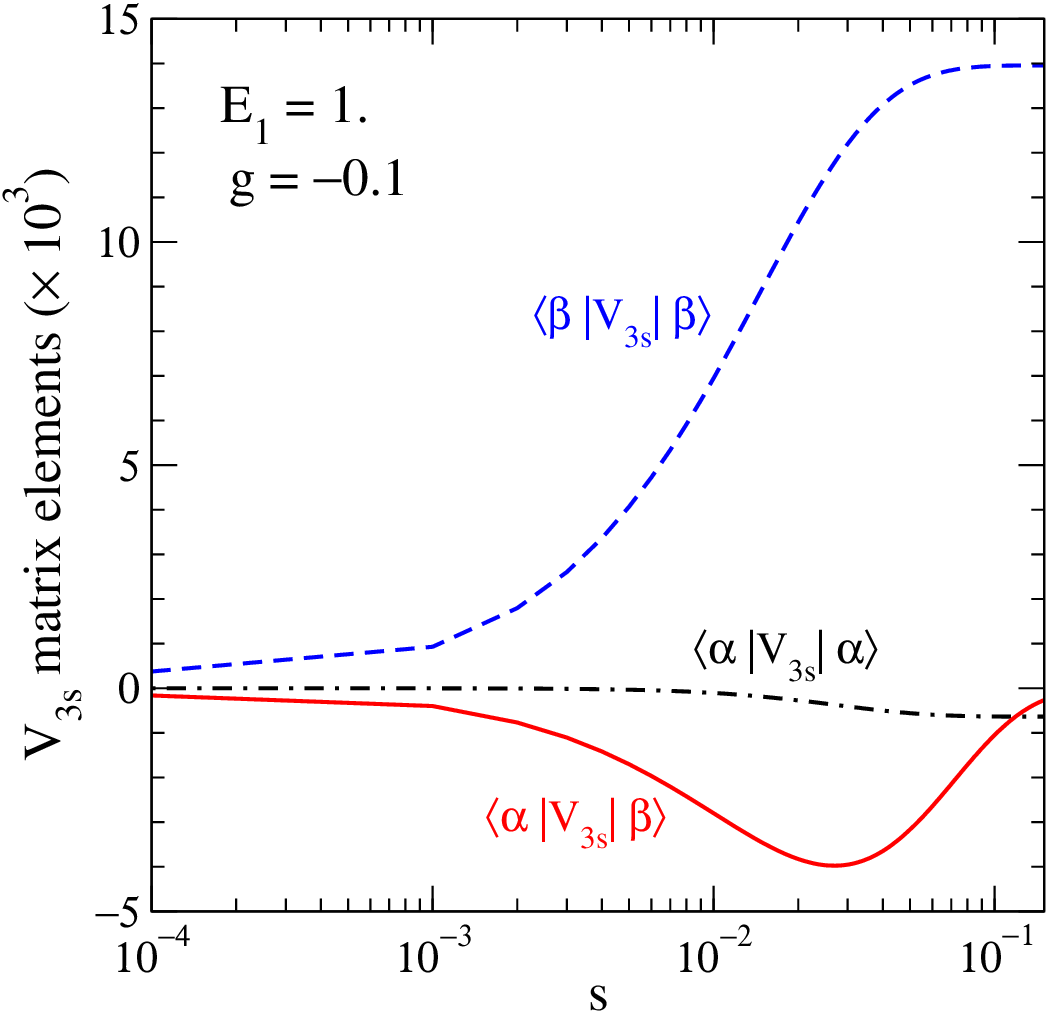}
  \hspace*{.1in}
  \includegraphics*[width=3.1in]{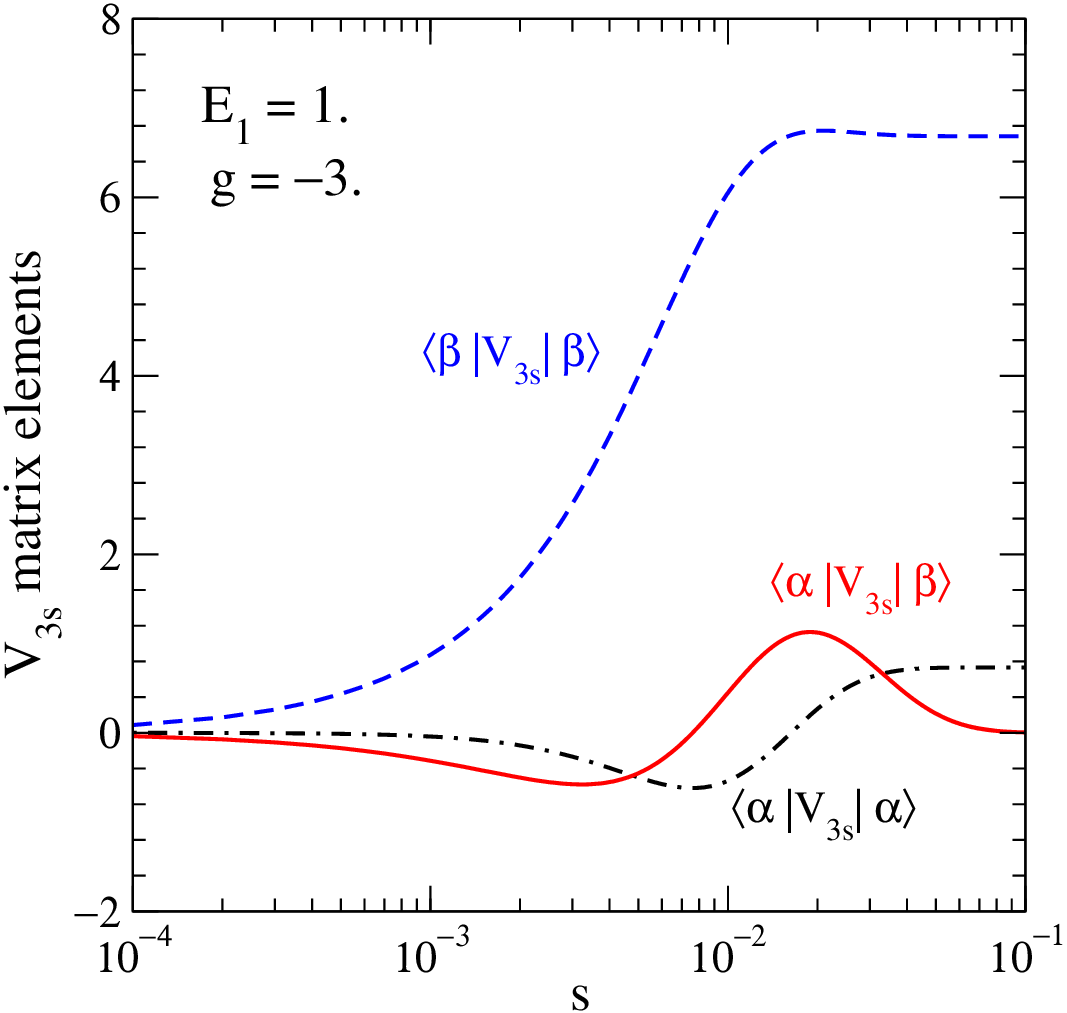}
  \caption{Hamiltonian matrix elements of $V_3$ as a function of the flow
  parameter $s$ in the three-boson sector for weak and strong couplings.}
  \label{fig:v3matrix}
\end{figure*}

For an explicit demonstration problem that can be exactly solved and readily
extended, we choose a zero-range initial interaction:
\beqn
V_{2}(x_i,x_j)=2 L g \, \delta(x_i-x_j) \;.
\eeqn
Note that for $g > 0$ this is repulsive. 
Matrix elements of $V_{2}$ between unsymmetrized states are:
\bea
(n_1 n_2 | V_{2} | n_3 n_4) &=& g \bigl[
   \delta_{0,n_1-n_2+n_3-n_4} + \delta_{0,n_1-n_2-n_3+n_4} \nonumber
   \\[2mm]
 &&  \qquad\null -\delta_{0,n_1-n_2+n_3+n_4} - \delta_{0,n_1-n_2-n_3-n_4} 
   - \delta_{0,n_1+n_2+n_3-n_4}  
 \nonumber \\[2mm]
  &&  \qquad\null -\delta_{0,n_1+n_2-n_3+n_4}   +
  \delta_{0,n_1+n_2-n_3-n_4} \bigr]  \;.
\eea
From these we readily compute the initial interaction matrix elements for
symmetrized states, which are given in Table~\ref{tab:5}.

In figs.~\ref{fig:v2matrix} and \ref{fig:v3matrix} we plot two-boson
matrix elements of $V_{2 \flow}$ and three-boson matrix elements of
$V_{3 \flow}$, for weak and strong couplings. We contrast flow for
$g=-0.1$ and $g=-3.0$, showing unitary renormalization group evolution
as a function of $s$. The energy scale is set by choosing $E_1=1$. 

The trace of $V_2$ in the two-boson sector is fixed, so we always see
one diagonal matrix element increase and the other decrease. Since the
off-diagonal matrix element must die exponentially, it is not
surprising to find similarity in fig.~\ref{fig:v2matrix} of how $V_2$
flows for strong and weak coupling. For $g=-0.1$ the rate of
exponential convergence is controlled by the linear term on the
right-hand-side of eq.~\ref{eq:vab}, while for $g=-3.0$ the
second-order term contributes significantly. This will be detailed in
the next section. To compare this flow with that of a $2\times 2$
matrix, note that here $\Omega=-3 E_1$ and for weak coupling the
exponent will be approximately $36 s$, which is why we see convergence
by the time $s \approx 0.1$. Convergence improves as the coupling
increases, but the range of $s$ over which convergence is seen will not
change drastically until $g$ is orders of magnitude larger than $E_1$.

$V_3$ displays more interesting behavior. Its trace in the three-boson
sector is not fixed, only the trace of $V_2+V_3$ is fixed. The trace of
$V_2$ in the three-boson sector varies because of its evolution in the
two-boson sector, so the trace of $V_3$ varies to restore unitarity.
Here we see $\wbb$ becoming much greater than $\waa$ for both weak and
strong coupling, because $\wbb$ is fed by a term second-order in $V_2$,
which we will identify with a tree diagram in the next section. 

For $g=-0.1$, $\wbb$ grows to ${\cal O}(g^2)$ while $\waa$ only grows
to  ${\cal O}(g^3)$. $\wab$ is initially comparable to $\wbb$ because
it is also fed by an ${\cal O}(V_2^2)$ tree-level interaction, but it
is then driven exponentially to zero with all off-diagonal matrix
elements. Additional features of this flow will be discussed in the
next section.

\begin{figure}[htb]
\includegraphics*[width=3.2in]{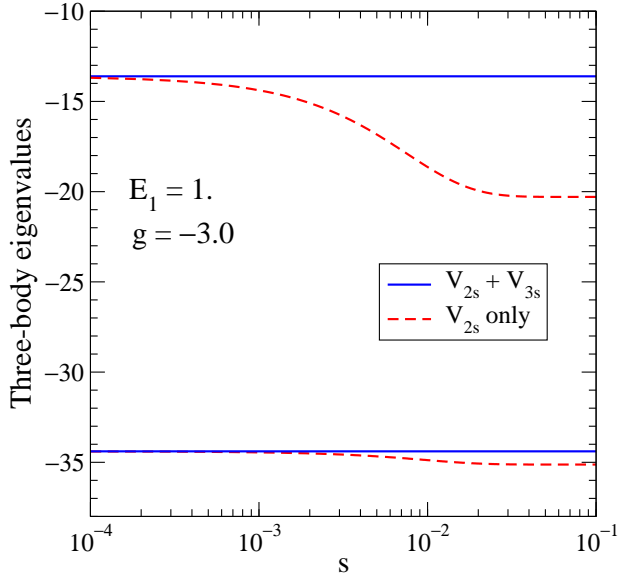}
\caption{The two lowest 
  eigenvalues of the three-boson systems as a function of $s$ 
  with the full interaction and when the three-body 
       interaction is omitted.}
\label{fig:no3error}
\end{figure}

One measure of the importance of $V_{3 \flow}$ is the relative error in the
three-boson eigenvalues when $V_{3 \flow}$ is not included; the errors arise
because $V_{2 \flow}$ alone can not produce unitary flow in the three-boson
sector.  The eigenvalues for these two cases using $g=-3.0$ are shown in
fig.~\ref{fig:no3error}. The eigenvalues are constant when the full interaction
is included (``It's unitary!''), while omitting $V_{3 \flow}$ leads to a small
error for the ground state and a 50\% error for the excited state. These errors
are consistent with the missing matrix elements, $\waa$ and $\wbb$, shown in
fig.~\ref{fig:v3matrix}. For $g=-0.1$ the errors are too small to be easily
visible.

Of course, there is no reason to restrict the initial hamiltonian to include
two-body interactions alone. The unitary flow of $V_{2 \flow}$ is not altered if
we add an initial three-body interaction, but that of $V_{3 \flow}$ can be
altered drastically. None of the flow equations derived above change, only the
initial values of $V_3$ matrix elements change. Off-diagonal matrix elements are
still driven to zero and diagonal matrix elements yield the correct eigenvalues.

We illustrate how a three-body force added from the start
affects unitary renormalization group evolution using
another zero-range initial interaction:
\beqn
V_{3}(x_i,x_j,x_k)=(2 L)^2 g_3 \, \delta(x_i-x_j) \, \delta(x_j-x_k)\;.
\eeqn
The initial matrix elements we need are $\waa=10 g_3$, $\wbb=6 g_3$ and $\wab=5 g_3$.
Note the relatively large prefactors.

\begin{figure*}[thb]
  \includegraphics*[width=3.05in]{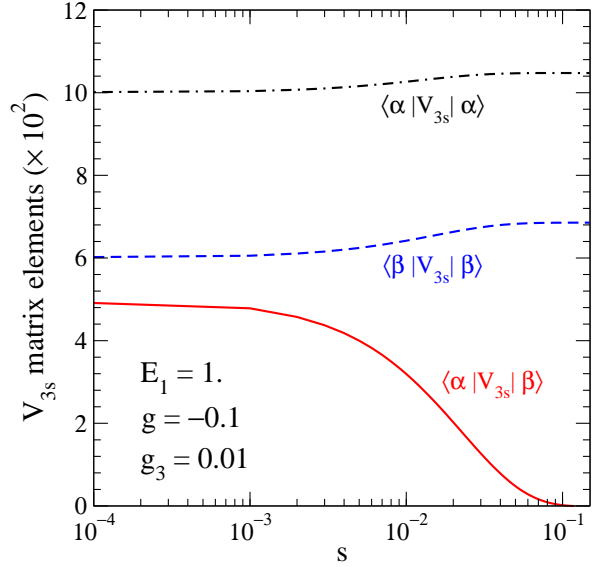}
  \caption{Hamiltonian matrix elements of $V_3$ as a function of the flow
  parameter $s$ in the three-boson sector for weak coupling, $g=-0.1$ and $g_3=0.01$.}
  \label{fig:g3matrixpert}
\end{figure*}

In fig.~\ref{fig:g3matrixpert} we display typical perturbative evolution, choosing
$g=-0.1$ and $g_3=0.01$ so that $V_2$ and $V_3$ have comparable perturbative effects. The
dominant effect of the perturbative three-body force is to both shift and split the
three-boson energies. The off-diagonal matrix element decays exponentially to zero with
little additional effect on the energies.

\begin{figure*}[thb]
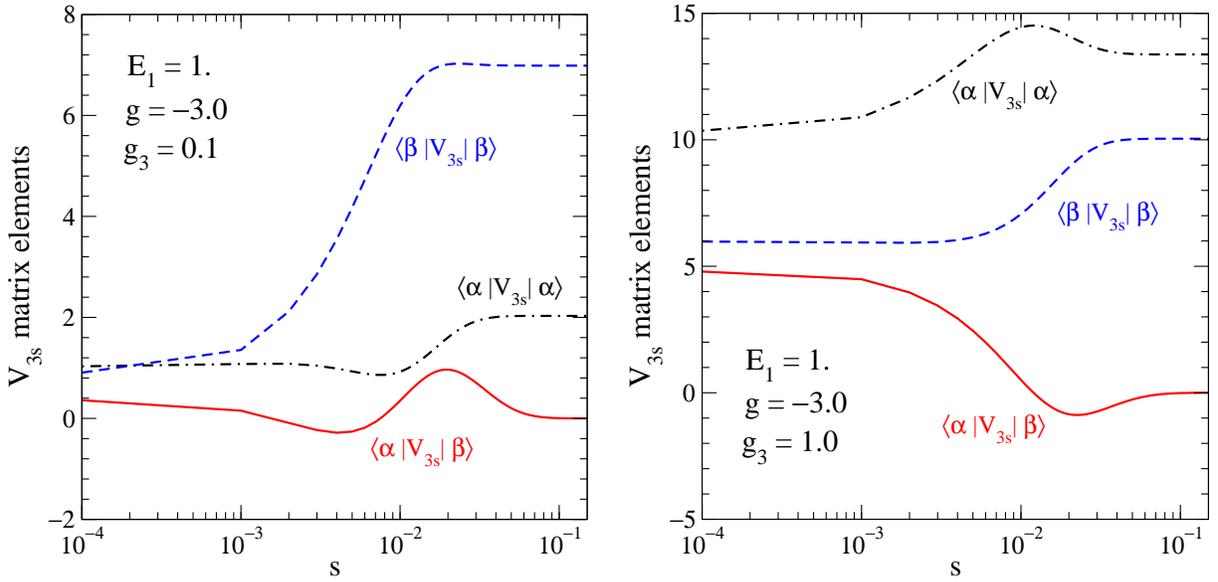

  \includegraphics*[width=3.05in]{g3sol1}
  \hspace*{.1in}
  \includegraphics*[width=3.05in]{g3sol2}
  \caption{Hamiltonian matrix elements of $V_3$ as a function of the flow
  parameter $s$ in the three-boson sector for two sets of strong couplings.}
  \label{fig:g3matrix}
\end{figure*}

\begin{figure*}[thb]
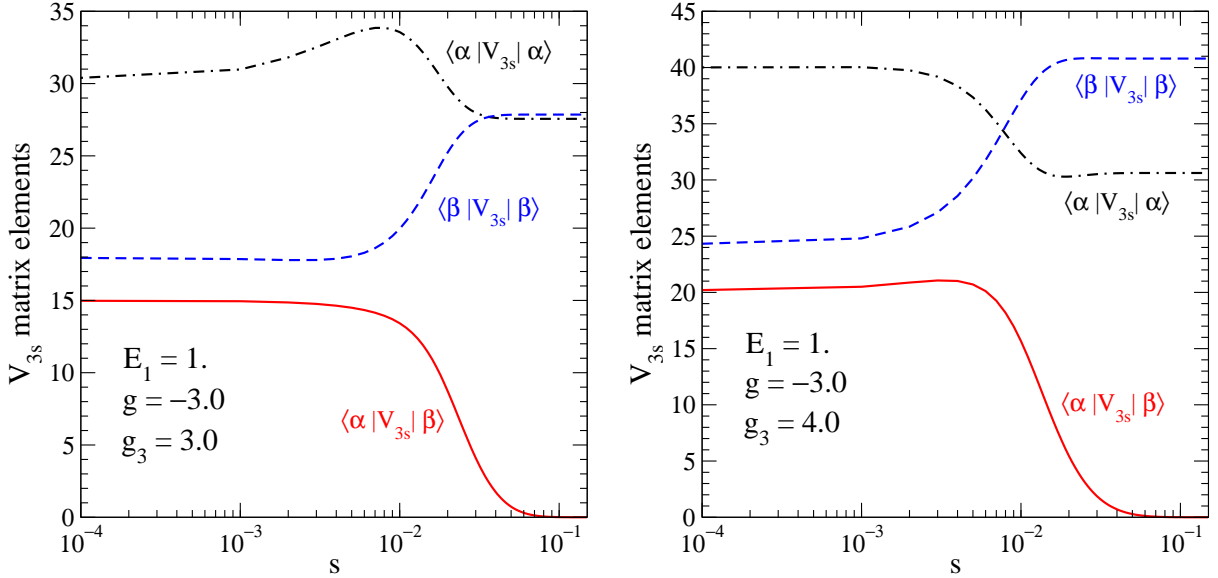

  \includegraphics*[width=3.05in]{g3sol3}
   \hspace*{.1in}
  \includegraphics*[width=3.05in]{g3sol4}
  \caption{Hamiltonian matrix elements of $V_3$ as a function of the flow
  parameter $s$ in the three-boson sector for two additional sets of strong couplings.}
  \label{fig:g3matrix2}
\end{figure*}

In fig.~\ref{fig:g3matrix} we display two examples of non-perturbative
evolution, both with $g=-3.0$ as considered above, with $g_3=0.1$ an example
where $V_2$ dominates the evolution and $g_3=1.0$ an example where $V_3$
dominates in the three-boson sector. These should be contrasted with fig.
\ref{fig:v3matrix} in which $g_3=0$. Perhaps the most interesting feature to
emerge as $g_3$ increases is that $V_3$ splits the three-boson energies with a
different sign for intermediate values of $g_3$. 
Figure~\ref{fig:g3matrix2} shows
that as $g_3$ is further increased, the order of its three-boson splittings
reverts to the perturbative ordering. This simple model actually displays a
remarkably broad range of interesting, analytically understandable behavior as a
function of the two couplings and our examples are not exhaustive.


\section{Fock Space Diagrammatic Analysis}
\label{sec:diagrams}

While it is straightforward to generate coupled evolution equations for
many-body interactions by computing explicit matrix elements of the operator
evolution equations, this requires inserting a complete set of $N$-boson states
to determine $V_N$ and does not immediately expose disconnected interactions in
which spectators do not participate. If one seeks a numerically tractable
differential equation for the evolution of $V_N$ in momentum representation,
these disconnected matrix elements are best avoided because they contain
momentum delta-functions for the spectator states. One of the  advantages of a
diagrammatic analysis is that it isolates these matrix elements as disconnected
diagrams, which are exactly cancelled as long as $V_{N-1}$ is fixed by
solving the SRG equations in sectors with fewer than $N$ particles.

We want to develop simple diagrammatic representations of
$[[\Hzero,V],\Hzero+V]$, to generate explicit unitary flow equations for matrix
elements of $V$. We must choose which many-boson matrix elements to represent.
The matrix elements that immediately lead to simple diagrammatic rules are
symmetrized, but not normalized for states containing more than one boson in a
given mode. Since we use the same states on both sides of the unitary flow
equation, normalization does not matter until we compute the explicit
hamiltonian.

Eigenstates of $\Hzero$ define particle
creation and annihilation operators, $| n \rangle=\ad_n | 0 \rangle$,
with $[a_m,\ad_n]=\delta_{mn}$, in terms of which,
\beqn
| i, j, k, ... \} \equiv \ad_i \ad_j \ad_k  \cdot \cdot \cdot | 0 \rangle \;,
\eeqn
where $i$, $j$, $k$, \ldots label external and internal legs in the diagrams. We
associate an eigenvalue of $\Hzero$, $E_i$, with any line labelled by
$i$. For two modes,
\beqn
\Hzero =E_1 \ad_1 a_1 + E_2 \ad_2 a_2 \;,
 \eeqn
 and for any number of modes,
 \beqn
 \Hzero = \sum_i E_i \ad_i a_i \;.
 \eeqn
 $V_2$ is then specified by $\{ij | V_2 | kl\}$ and $V_3$ is specified
by $\{ijk | V_3 | lmn \}$: 
 \beqn
 V_2 = \Bigl(\frac{1}{2!}\Bigr)^2~\sum_{ijkl} \ad_i \ad_j \{ij | V_2 | kl\} a_l a_k  \;,
 \eeqn
 \beqn
 V_3 = \Bigl(\frac{1}{3!}\Bigr)^2~\sum_{ijklmn} 
   \ad_i \ad_j \ad_k \{ijk | V_3 | lmn\} a_n a_m a_l \;.
 \eeqn
In our simple example above, these indices
took on only the values 1 and 2, but this constraint does not simplify
the diagrammatic rules and we need impose it only at the end to verify
that the rules produce the same equations as derived in the previous section.

\begin{figure*}[htb]
  \includegraphics*[width=6.2in]{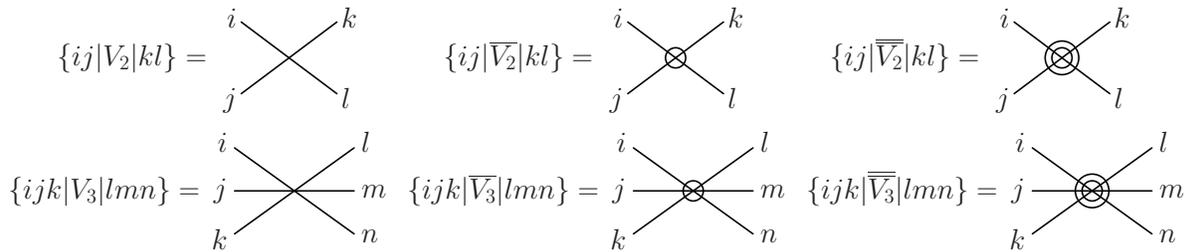}
  \caption{Labelled vertices represent matrix elements of  $V_N$,
  $\mybar{V}_N \equiv [T,V_N]$, and $\mybar{\mybar{V}}_N \equiv [[T,V_N],T]$.}
  \label{fig:srgdiagrams1}
\end{figure*}

We are now in a position to develop a complete set of diagrammatic
rules for the right-hand-side of the flow equations for $\{ij | V_2 |
kl\}$ and $\{ijk | V_3 | lmn \}$. In fig.~\ref{fig:srgdiagrams1} we
show a diagrammatic representation for the basic elements of these flow
equations. All terms are built from $V_N$, $\mybar{V}_N \equiv [T,V_N]$ and
$\mybar{\mybar{V}}_N \equiv [[T,V_N],T]$, where $\{ij | \mybar{V}_2 | kl\}
=(E_i+E_j-E_k-E_l) \{ij | V_2 | kl\}$, and similar relations are easily
found for all matrix elements of $\mybar{V_N}$ and $\mybar{\mybar{V_N}}$. External
legs on these diagrams are labeled with symmetrized labels, so we will
not count diagrams as ``topologically distinct"  if they result from a
permutation of indices associated with the legs of a single vertex. The
redundancies from these permutations cancel the $(1/2!)^2$ in
$V_2$ and the $(1/3!)^2$ in $V_3$, except in diagrams with loops, where an $N$-loop
diagram is weighted by $1/(N+1)!$. Diagrams with one loop are assigned a factor of $1/2$,
etc.

\begin{figure*}[t]
  \includegraphics*[width=5.5in]{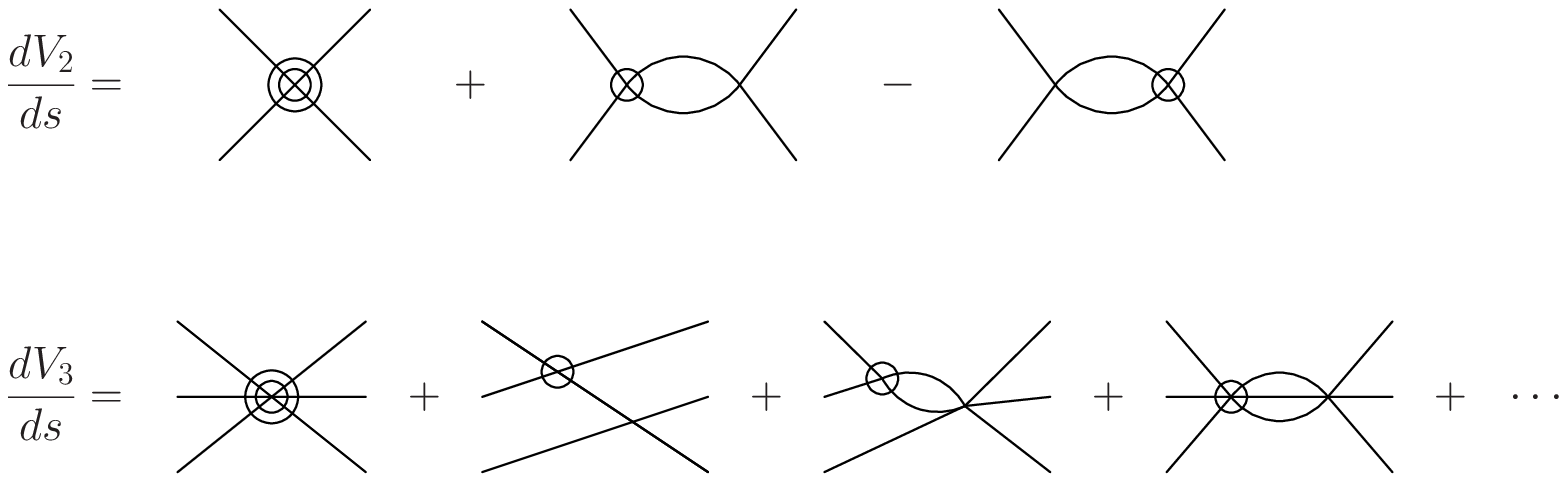}
  \caption{Complete diagrammatic representation of $dV_2/ds$ and a
  schematic representation of the diagrams for $dV_3/ds$.
  For the latter, there are eighteen tree diagrams,
six one-loop diagrams, and two two-loop diagrams.}
  \label{fig:srgdiagrams2}
\end{figure*}

In fig.~\ref{fig:srgdiagrams2} we show a schematic representation of the
diagrams that appear in the SRG flow equations in the
two-boson and three-boson sectors. Diagrammatic rules
are easily deduced from the equations and are readily generalized to any
sector. For clarity we expand 
eq.~(\ref{eq:commutatorboth}):
\bea
  \frac{dV_{2 \flow}}{d\flow} +\frac{dV_{3 \flow}}{d\flow} 
  &=&  [ [\Hzero, V_{2 \flow}], \Hzero]+[ [\Hzero, V_{3 \flow}], \Hzero]
  \nonumber \\[2mm]
  &&  \null +[[\Hzero, V_{2 \flow}],V_{2 \flow}]+
  [[\Hzero, V_{2 \flow}],V_{3 \flow}]+[[\Hzero, V_{3 \flow}],V_{2 \flow}]
     +[[\Hzero, V_{3 \flow}],V_{3 \flow}]
  \nonumber \\[2mm]
  &=&  \mybar{\mybar{V}}_2+\mybar{\mybar{V}}_3+\Bigl\{ 
    \Bigl([\mybar{V}_2,V_2]+[\mybar{V}_2,V_3]+[\mybar{V}_3,V_2]
+[\mybar{V}_3,V_3]\Bigr)  \nonumber \\[2mm]
  && \qquad\qquad - \Bigl(V_2 \leftrightarrow \mybar{V}_2, V_3 
    \leftrightarrow \mybar{V}_3] 
  \Bigr) \Bigr\}
  \;.
  \label{eq:commutator2}
\eea 
Diagrams represent symmetrized matrix elements of these equations. Every line
carries a label and products of $\mybar{V}$ and $V$ result in contractions that
equate indices on lines between the two vertices. Once indices are assigned to
all lines, vertices represent matrix elements of $V_2$, $V_3$, $\mybar{V}_2$
and $\mybar{V}_3$, with indices matched to lines. There is a sum over all
internal indices, so even tree diagrams represent sums. One-loop diagrams
appear in the evolution of $V_2$, while both one- and two-loop diagrams govern
the evolution of $V_3$. In general, tree diagrams through $(N-1)$-loop diagrams appear in the
evolution of $V_N$.

The structure of eq.~(\ref{eq:commutator2})  guarantees that $V_{2
\flow}$ is completely determined by two-boson matrix elements, with
$V_{3 \flow}$ then determined by three-boson matrix elements. Comparing
eq.~(\ref{eq:commutator2}) with fig.~\ref{fig:srgdiagrams2}, note that
$V_2$ and $V_3$ appear together in the equation but there are separate
diagrammatic equations for their evolution. Two-boson matrix elements
of eq.~(\ref{eq:commutator2}) produce the diagrammatic equation for
$dV_2/ds$, while three-boson matrix elements produce the
diagrammatic equation for $dV_3/ds$. When three-boson matrix
elements of eq.~(\ref{eq:commutator2})  are computed, the matrix
elements of $dV_2/ds$ on the left-hand-side cancel against
diagrams on the right-hand-side that are identical to the two-boson
diagrams with a spectator line that is not connected to either vertex.
As long as $dV_2/ds$ solves the two-boson evolution equations,
these disconnected diagrams can be dropped on the right-hand-side of
the diagrammatic equation and only $dV_3/ds$ appears on the
left-hand-side.

There are two reasons that disconnected diagrams never appear. Diagrams
in which both vertices are not connected to one another by at least one
line vanish because of the commutator structure, $[\mybar{V},V]$. If
there are no contractions between creation and annihilation operators
in these two vertices, the two terms from the commutator produce
identical matrix elements that exactly cancel.  

As discussed above, diagrams in which an external line is not connected
to any vertex, such as three-boson diagrams in which a single spectator
line appears with the one-loop interactions between the other two
bosons as displayed below in fig.~\ref{fig:srgdiagrams3}, simply
produce copies of the flow equations from lower sectors of Fock space.
So, if $V_2$ satisfies the two-boson flow equations, disconnected
three-boson matrix elements of $dV_2/ds$ cancel the
disconnected matrix elements of $[\mybar{V_2},V_2]$. We isolate the
flow equation for $dV_3/ds$ simply by dropping these
disconnected diagrams and setting $dV_3/ds$ equal to the sum of
connected diagrams, including the diagram for $\mybar{\mybar{V_3}}$.

So, $V_2$ is determined by two-boson matrix elements, because $V_3$,
$V_4$, etc. will not appear in this sector and all diagrams that do
appear in the two-boson sector also appear in the three-boson sector
with one disconnected line, in the four-boson sector with two
disconnected lines, etc. Again, if $V_2$ satisfies the flow equation in
the two-boson sector, no disconnected diagrams survive in the flow
equation for $V_3$ in the three-boson sector. In the three-boson
sector, there are tree diagrams involving two powers of $V_2$, one-loop
diagrams involving one power of $V_2$ and one power of $V_3$, and
two-loop diagrams involving two powers of $V_3$. $V_2$ is already
determined by the two-boson evolution equation and $V_3$ is the only
new interaction appearing in the three-boson evolution equations, so
the three-boson evolution equations determine $V_3$. 

This type of analysis is easily generalized to higher sectors because
of the simple second-order structure of the exact unitary flow
equation. $V_4$ is determined by the four-boson evolution equations.
These include tree-diagrams with one power or $V_2$ and one of $V_3$,
one-loop diagrams with one power of $V_2$ and one power of $V_4$ or
with two powers of $V_3$,  two-loop diagrams with one power of $V_3$
and one power of $V_4$, and three-loop diagrams that are second-order
in $V_4$. 

\begin{figure*}[htb]
  \includegraphics*[width=5in]{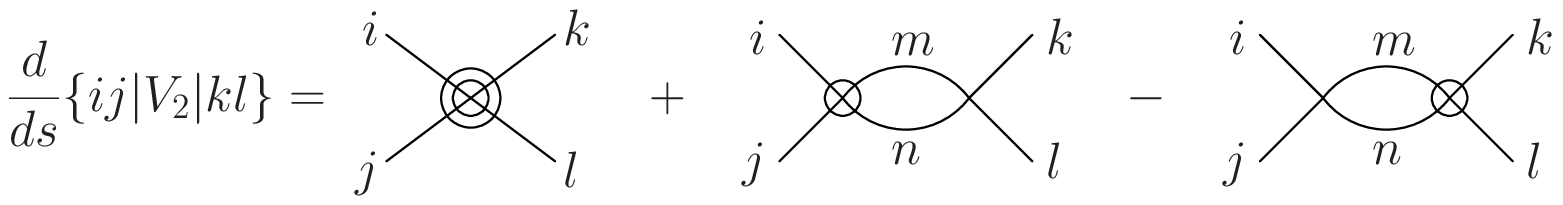}
  \caption{Diagrammatic representation of eq.~(\ref{eq:dV2ds}).}
  \label{fig:srgdiagrams3}
\end{figure*}

In fig.~\ref{fig:srgdiagrams3} the complete set of labelled
two-boson diagrams is
shown, from which we obtain
\beqn
\frac{d}{ds} \{ij | V_2 | kl \} = 
  \{ij | \mybar{\mybar{V}}_2 | kl \} + \frac{1}{2!} \sum_{mn} 
  \bigl( \{ij | \mybar{V}_2 | mn \}  
  \{mn | V_2 | kl \} - \{ij | V_2 | mn \}  
  \{mn | \mybar{V}_2 | kl \} \bigr) \;.
  \label{eq:dV2ds}
\eeqn
Substituting for $\mybar{V}_2$ and $\mybar{\mybar{V}}_2$ we find
\beqn
\frac{d}{ds} \{ij | V_2 | kl \} = -\bigl(E_{ij}-E_{kl}\bigr)^2
   \{ij | V_2 | kl \} + \frac{1}{2!} \sum_{mn} \bigl( E_{ij}+E_{kl}-2 E_{mn} \bigr) 
   \{ij | V_2 | mn \}  \{mn | V_2 | kl \} \;,
   \label{eq:dV2dsp}
\eeqn
where $E_{ij}=E_i+E_j$. It is straightforward to show that when sums are
restricted to the lowest two modes,  eqs.~(\ref{eq:vaa})--(\ref{eq:vab}) are
reproduced.

\begin{figure*}[htb]
  \includegraphics*[width=5in]{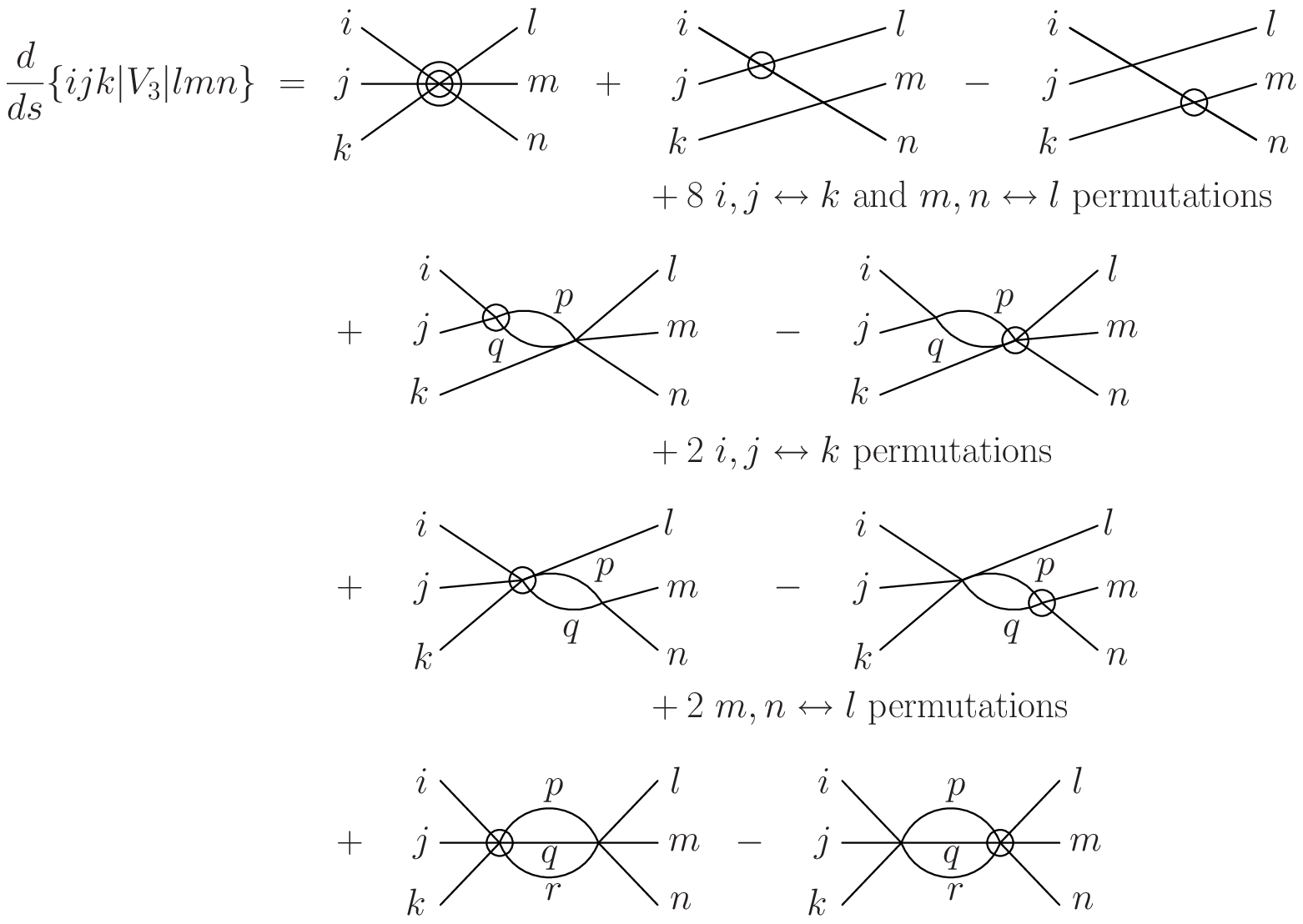}
  \caption{Diagrammatic representation of eq.~(\ref{eq:dV3ds}).
  The notation $i,j\leftrightarrow k$ means $i\leftrightarrow j$
  or $j\leftrightarrow k$.}
  \label{fig:srgdiagrams4}
\end{figure*}

In fig.~\ref{fig:srgdiagrams4} the complete set of labelled
three-boson diagrams is
shown. This equation is for the evolution of $V_3$, with $V_2$ solving the
two-boson equations above and no disconnected diagrams appearing as a result.
There are eighteen distinct tree-level diagrams, twelve one-loop diagrams and two
two-loop diagrams. 
We do not explicitly display all possible external
legs in the diagrams or the equations but list them as permutations.
The resulting equation is
\bea
\frac{d}{ds} \{ijk | V_3 | lmn \} &=& \{ijk | \mybar{\mybar{V}}_3 | lmn \} 
  \nonumber \\[1mm]
&&  \hspace*{-.8in}
+ \sum_{p} \bigl( \{ij | \mybar{V}_2 | lp \}  \{pk | V_2 | mn \} - \{ij | V_2 | lp \}  
   \{pk | \mybar{V}_2 | mn \} 
  \nonumber \\[0mm]
&& \hspace*{.2in}
  \null + 8~ i,j \leftrightarrow k, ~m,n \leftrightarrow l ~\mbox{permutations} \bigr)
  \nonumber \\[1mm]
&&  \hspace*{-.8in}
  +\frac{1}{2!} \sum_{pq} \bigl( \{ij | \mybar{V}_2 | pq \}  \{pqk | V_3 | lmn \} -
\{ij | V_2 | pq \}  \{pqk | \mybar{V}_3 | lmn \} 
  \nonumber \\[0mm]
&& \hspace*{.2in}
  \null + 2~ i,j \leftrightarrow k ~ \mbox{permutations} \bigr)
  \nonumber \\[1mm]
&&  \hspace*{-.8in}
  +\frac{1}{2!} \sum_{pq} \bigl( \{ijk | \mybar{V}_3 | lpq \}  \{pq | V_2 | mn \} -
\{ijk | V_3 | lpq \}  \{pq | \mybar{V}_2 | mn \} 
  \nonumber \\[0mm]
&& \hspace*{.2in}
  \null + 2 ~m,n \leftrightarrow l ~ \mbox{permutations} \bigr)
  \nonumber \\[1mm]
&&  \hspace*{-.8in}
  +\frac{1}{3!} \sum_{pqr} \bigl( \{ijk | \mybar{V}_3 | pqr \}  
    \{pqr | V_3 | lmn \} -
   \{ijk | V_3 | pqr \}  \{pqr | \mybar{V}_3 | lmn \} \bigr)     
   \;. 
     \label{eq:dV3ds}
\eea
Most of the terms on the right-hand-sides of these equations are given
by permutations of the indices in the terms explicitly displayed. For
example, in the tree diagrams in fig.~\ref{fig:srgdiagrams4}, one
particle line on the left is connected to a different vertex than the
other two. We must explicitly add each permutation of the index
assigned to this line, obtaining additional diagrams in which this
index is $i$ instead of $k$ and $j$ instead of $k$. Permutations of the
two lines connected to the same vertex are considered ``topologically
identical" to the diagram shown, so they produce no new diagrams. This
reflects the fact that $ \{ij | V_2 | pq \} =\{ji | V_2 | pq \}  $.

In front of each sum over $N$ indices there is a factor $1/N!$, so to each
$N$-loop diagram we associate this factor of $1/N!$. These sums are
unrestricted; if we replace them with sums over distinct
permutations the factor of $1/N!$ drops out.

It is again a straightforward but tedious exercise to show that
truncation to the two lowest modes yields
eqs.~(\ref{eq:waa})--(\ref{eq:wab}). To match equations, note that $| a
\rangle = \sqrt{1/2}~ | 11 \}$, $| b \rangle = \sqrt{1/2}~ | 22 \}$ and
$| c \rangle = | 12 \}$ are the two-boson states we need,  while $|
\alpha \rangle = \sqrt{1/6}~ | 111 \}$ and $| \beta \rangle =
\sqrt{1/2}~ | 122\}$ are the only positive parity three-boson states we
need.

If the hamiltonian has no three-body force at the start of SRG
evolution, one will be produced. Initially, the tree diagrams, which
are second-order in $V_2$, dominate. Next $\mybar{\mybar{V}}_3$ will
suppress far off-diagonal matrix elements while terms of order $V_2
V_3$ will be more important near the diagonal, with the two-loop
diagram emerging last since it is order $V_3^2$. Naturally, if $V_3$ is
not zero at the start of the renormalization group evolution, the
relative magnitude of these contributions can change. The diagrams
provide a tool for analyzing various contributions to the flow
equations and seeking approximations, which might be essential in more
realistic calculations.

\begin{figure*}[thb]
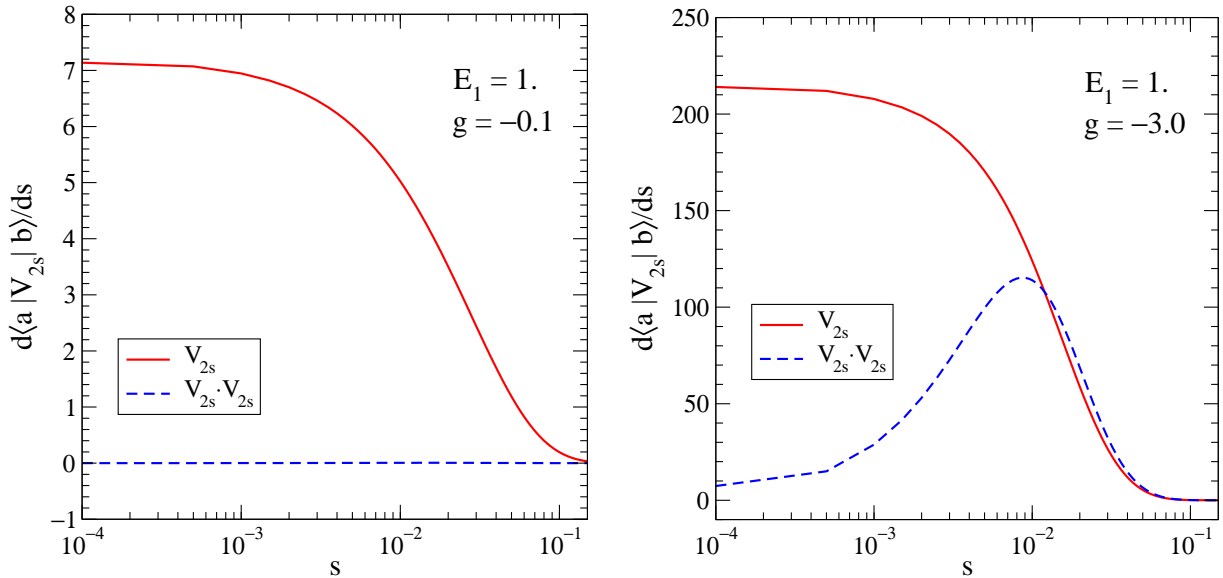

  \includegraphics*[width=3.05in]{v2ab_rhs_pert}
  \hspace*{.1in}
  \includegraphics*[width=3.10in]{v2ab_rhs_nonpert}
  \caption{Matrix elements that contribute to $\frac{d}{ds} \vab$ in
  the two-boson sector for weak and strong coupling. $\vab$ is dominated
  by $\mybar{\mybar{V_2}}$, with the second-order one-loop correction
  barely visible when $g=-0.1$.}
  \label{fig:v2ab_rhs}
\end{figure*}

\begin{figure*}[thb]
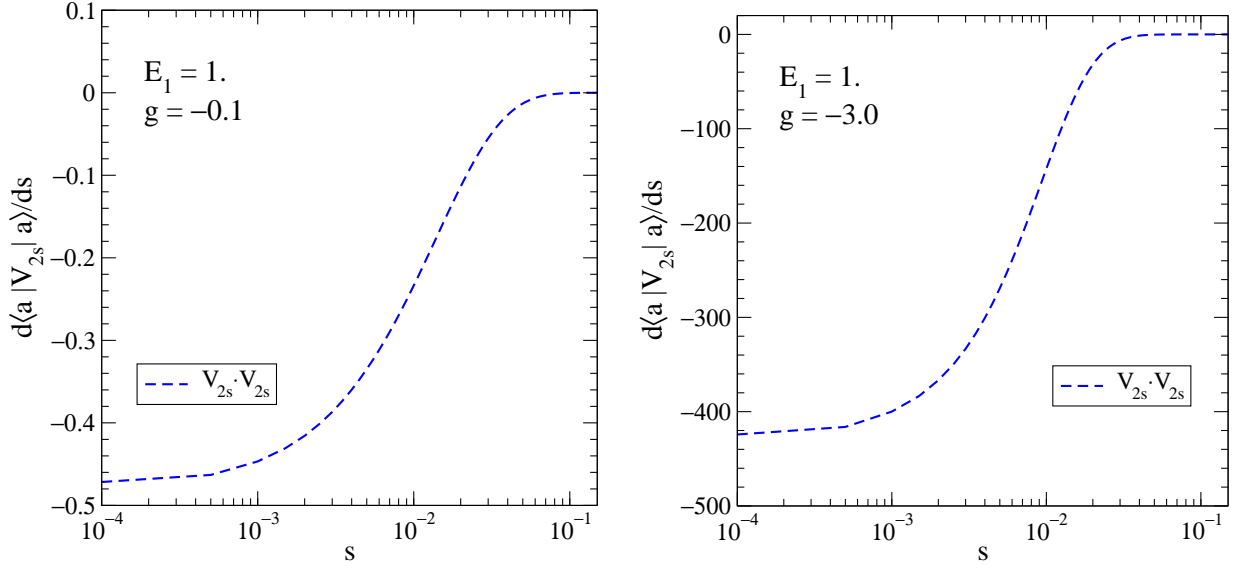

  \includegraphics*[width=3.1in]{v2aa_rhs_pert}
  \hspace*{.1in}
  \includegraphics*[width=3.1in]{v2aa_rhs_nonpert}
  \caption{Matrix elements that contribute to $\frac{d}{ds} \vaa$ in the two-boson sector
for weak and strong coupling. Only the one-loop diagram contributes.}
  \label{fig:v2aa_rhs}
\end{figure*}

\begin{figure*}[thb]
  \includegraphics*[width=3.1in]{v3ab_rhs_pert}
    \hspace*{.1in}
  \includegraphics*[width=3.1in]{v3ab_rhs_nonpert}
  \caption{Matrix elements that contribute to $\frac{d}{ds} \wab$ in the three-boson
sector for weak and strong coupling. Tree diagrams initially dominate the evolution of
$\wab$, allowing $\mybar{\mybar{V_3}}$ to grow and then drive $\wab$ to zero. Here again,
the one-loop and two-loop contributions are barely visible for $g=-0.1$ and the two-loop
contribution remains negligible even for strong coupling. This can change if $V_3$ is not
zero initially. }
  \label{fig:v3ab_rhs}
\end{figure*}

\begin{figure*}[thb]
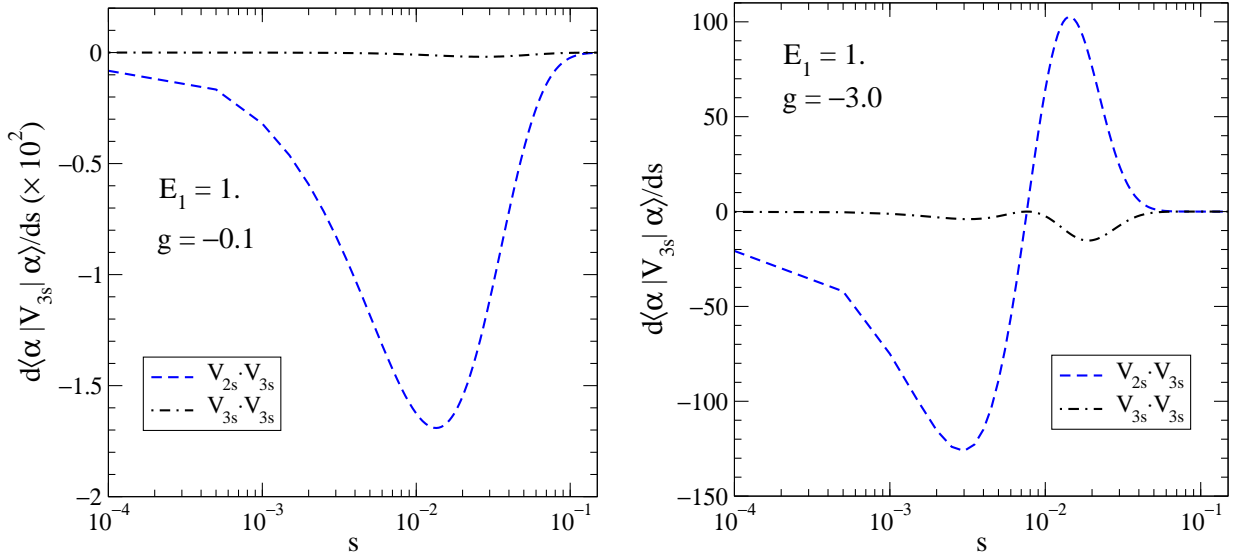

  \includegraphics*[width=3.1in]{v3aa_rhs_pert}
  \hspace*{.1in}
  \includegraphics*[width=3.1in]{v3aa_rhs_nonpert}
  \caption{Matrix elements that contribute to $\frac{d}{ds} \waa$ in the three-boson
sector for weak and strong coupling. No tree diagrams contribute to $\waa$, and the
relatively small one-loop contributions dominate over the two-loop contributions in both
cases. }
  \label{fig:v3aa_rhs}
\end{figure*}

\begin{figure*}[thb]
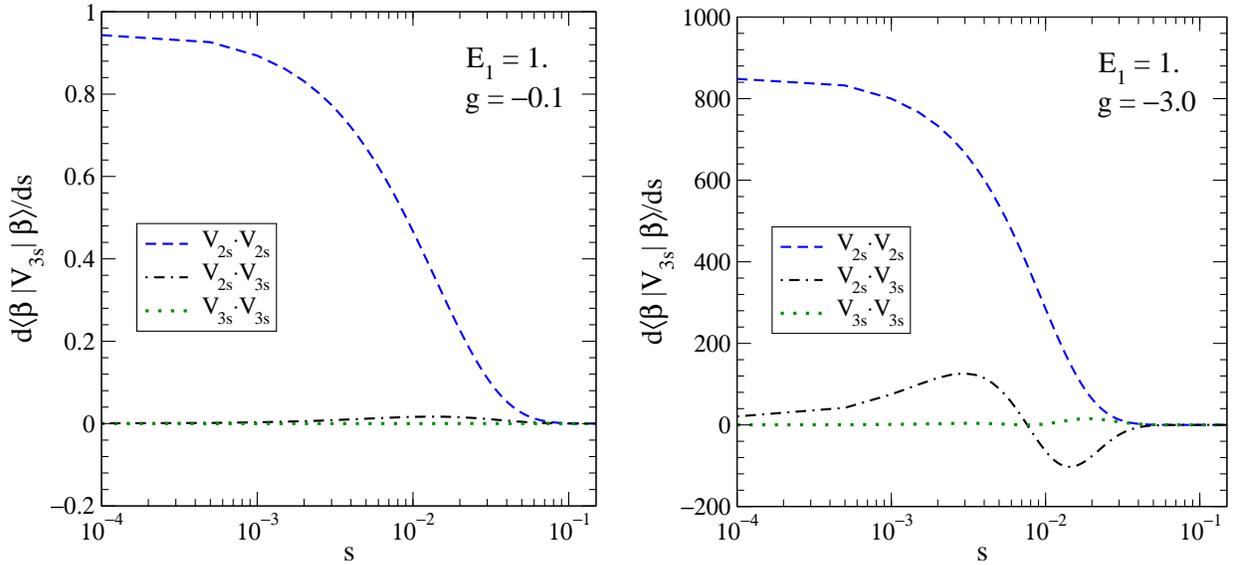

  \includegraphics*[width=3.1in]{v3bb_rhs_pert}
  \hspace*{.1in}
  \includegraphics*[width=3.1in]{v3bb_rhs_nonpert}
  \caption{Matrix elements that contribute to $\frac{d}{ds} \wbb$ in the three-boson
sector for weak and strong coupling. Tree diagrams dominate the initial evolution of
$\wbb$, with relatively small one-loop contributions and  negligible two-loop
contributions in both cases. }
  \label{fig:v3bb_rhs}
\end{figure*}

In figs.~\ref{fig:v2ab_rhs}--\ref{fig:v3bb_rhs} we contrast various
contributions to the evolution of $V_2$ and $V_3$ for weak coupling, $g=-0.1$,
and strong coupling, $g=-3.0$. Figure~\ref{fig:v2ab_rhs} displays contributions
to the off-diagonal matrix element $\vab$. In both cases the initial evolution
is dominated by the double-commutator, $\mybar{\mybar{V_2}}$, which always
drives off-diagonal matrix elements to zero exponentially. For weak coupling
the one-loop, ${\cal O}(V_2^2)$, contribution remains negligible throughout the
evolution, while for strong coupling it becomes comparable to that of
$\mybar{\mybar{V_2}}$ and helps drive $\vab$ to zero. This qualitative behavior
is what is seen when this transformation is applied to realistic
nucleon-nucleon interactions \cite{Bogner:2006srg,Bogner:2007jb}, but it is
possible to find cases where the one-loop contribution first opposes
$\mybar{\mybar{V_2}}$ and forces $\vab$ to grow before $\mybar{\mybar{V_2}}$
finally takes over \cite{Glazek:2007}. This happens when the running cutoff
encounters a bound state threshold and we make no attempt to discuss such
interesting exceptions in this paper, although they are inevitable in
three-body systems that display the Efimov effect \cite{Efi70,Efi71}.

Figure~\ref{fig:v2aa_rhs} shows the sole one-loop contribution to the
evolution of the diagonal matrix element $\vaa$. Since the trace of
$H_\flow$ is conserved in each sector of Fock space, contributions to
$\vbb$ simply have the opposite sign. The only change from weak to
strong coupling is the magnitude of the one-loop contribution. This
simplicity is partially a reflection of the drastic truncation of Fock
space but a similar exponential fall-off in strength is guaranteed in
more realistic examples as long as the transformation drives the matrix
to band-diagonal form. Not only do the far off-diagonal matrix elements
flow to zero exponentially, the diagonal matrix elements evolve rapidly
at first and then slow as they approach their asymptotic values. Again,
there are exceptions to this simple evolution if bound-state thresholds
are encountered \cite{Glazek:2007}.

In figs.~\ref{fig:v3ab_rhs}--\ref{fig:v3bb_rhs} we turn to terms that
contribute to the evolution of $V_3$. In fig.~\ref{fig:v3ab_rhs} we
immediately see that tree diagrams, which are ${\cal O}(V_2^2)$,
dominate the initial evolution of $\wab$ for both weak and strong
coupling. This is guaranteed when $V_3$ itself starts at zero, but the
subsequent evolution differs for weak and strong coupling. For weak
coupling, once $\wab$ itself has any appreciable strength,
$\mybar{\mybar{V_3}}$ builds and drives this off-diagonal matrix
element back to zero. One-loop and two-loop contributions remain
negligible for all $s$ when the coupling is weak, simply because these
terms are suppressed by powers of $g$. At strong coupling it is the
one-loop contribution that builds to shut off the growth of $\wab$ and
then drive it to zero, while $\mybar{\mybar{V_3}}$ and the two-loop
contribution remain small but non-negligible near the end of the
evolution.

\begin{figure*}[bt]
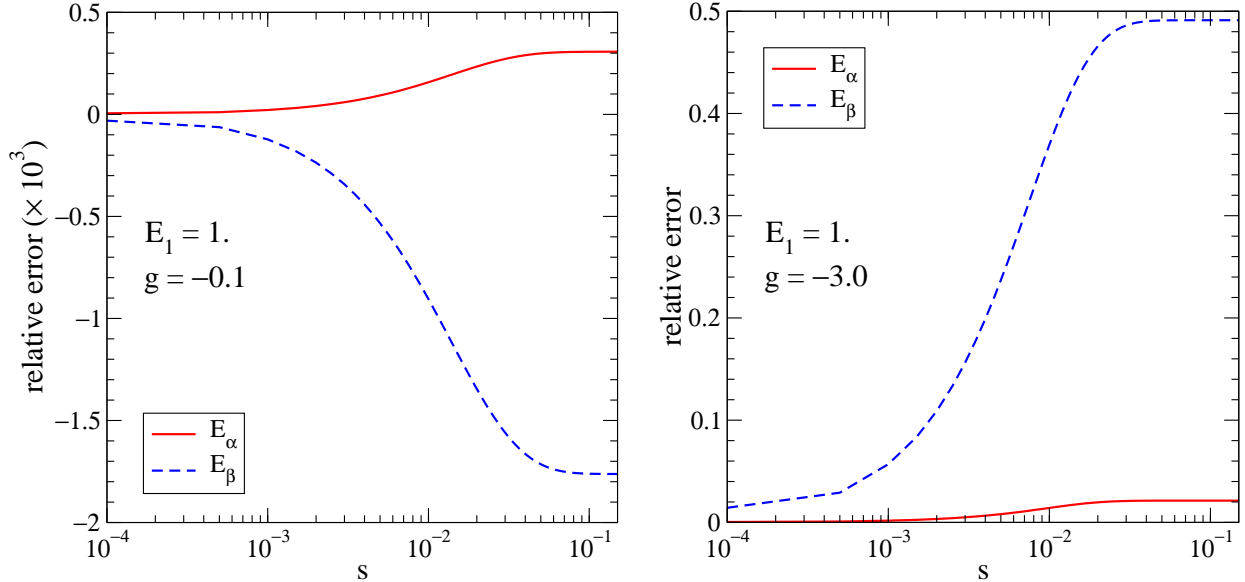

  \includegraphics*[width=3.2in]{no_v3_error_pert}
    \hspace*{.1in}
  \includegraphics*[width=3.05in]{no_v3_error_nonpert}
  \caption{Relative errors in the three-boson eigenvalues for weak and strong coupling
when $V_3$ is simply dropped. }
  \label{fig:no_v3_error}
\end{figure*}

The evolution of $\waa$ and $\wbb$ are not simply connected by a trace
constraint because $V_2$ also contributes to the trace of $H_\flow$ in
the three-boson sector. There is no tree diagram contributing to
$\waa$, so it remains small in comparison to $\wbb$ for both weak and
strong coupling. We see in fig.~\ref{fig:v3aa_rhs} that the one-loop
diagram dominates for both weak and strong coupling, but this
contribution changes sign during the evolution of $\waa$ for strong
coupling. Figure~\ref{fig:v3bb_rhs} shows that the tree diagram dominates
this evolution, with other contributions remaining negligible
throughout the evolution at weak coupling and only the one-loop
contribution becoming significant at strong coupling.

In fig.~\ref{fig:no_v3_error} we show the fractional error in the three-boson
eigenvalues $E_\alpha$ and $E_\beta$
when $V_3$ is simply dropped. This error remains smaller than
${\cal O}(g^2)$ for weak coupling, which is the order of the tree diagrams that
are ignored,  and grows to 50\% for strong coupling. In both cases it is 
$E_\beta$ that is most sensitive to $V_3$.

\begin{figure*}[tb]
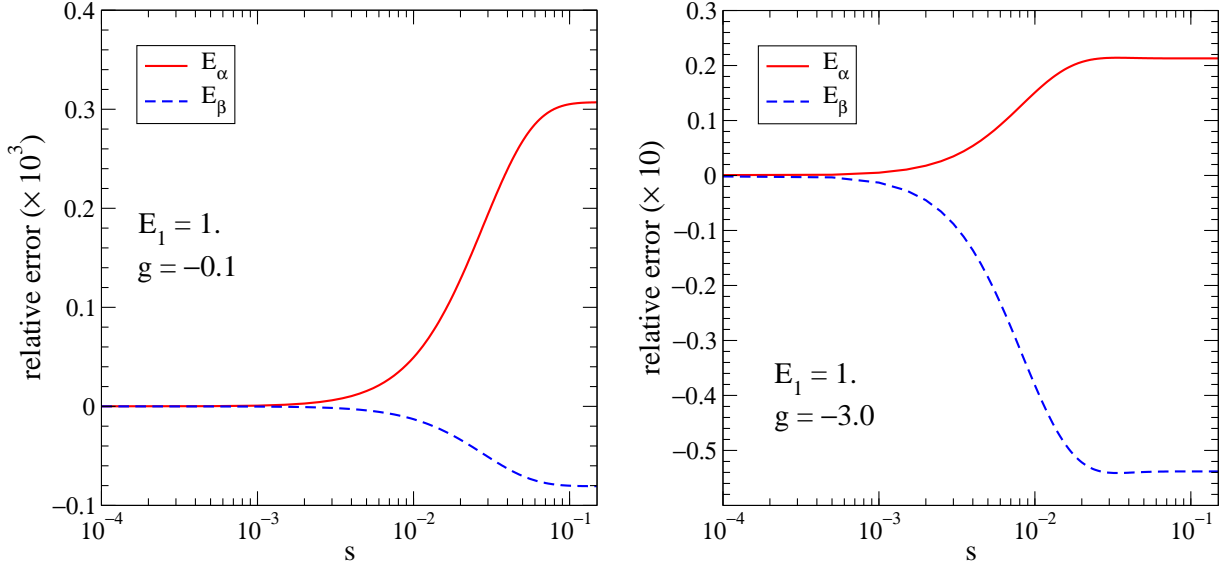

  \includegraphics*[width=3.1in]{tree_v3_error_pert}
  \hspace*{.1in}
  \includegraphics*[width=3.05in]{tree_v3_error_nonpert}
  \caption{Relative errors in the three-boson eigenvalues for weak and coupling when
$\mybar{\mybar{V_3}}$ and tree diagrams are used to evolve $V_3$.}
  \label{fig:tree_v3_error}
\end{figure*}

\begin{figure*}[tb]
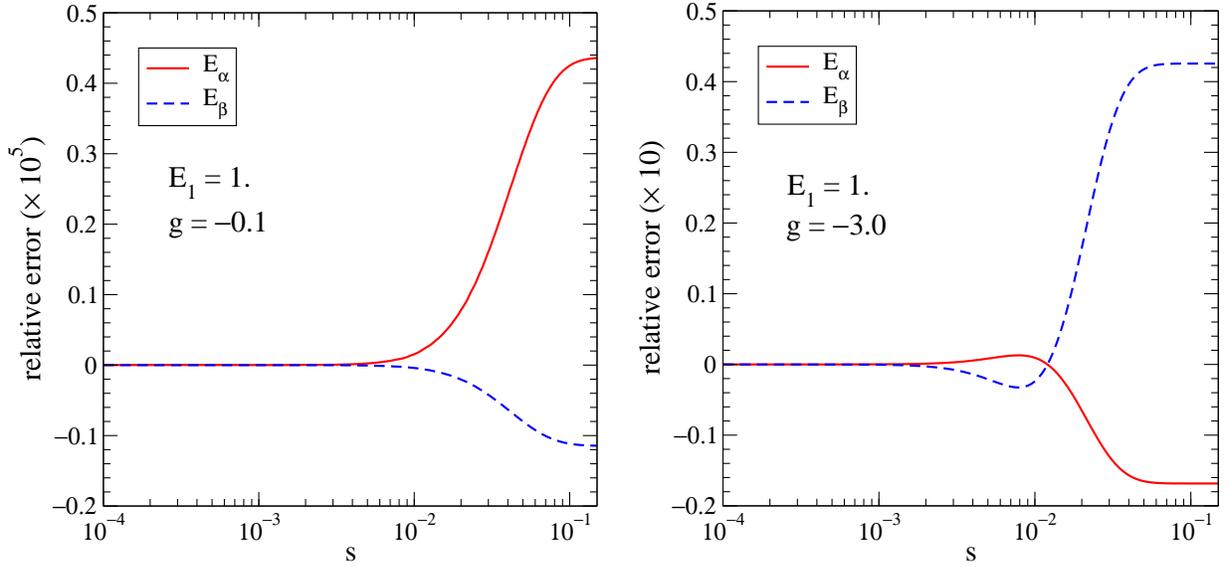

  \includegraphics*[width=3.1in]{1loop_v3_error_pert}
  \hspace*{.1in}
  \includegraphics*[width=3.05in]{1loop_v3_error_nonpert}
  \caption{Relative errors in the three-boson eigenvalues for weak and strong coupling
when tree and one-loop diagrams are used to evolve $V_3$.}
  \label{fig:1loop_v3_error}
\end{figure*}

In fig.~\ref{fig:tree_v3_error} we show these same 
fractional errors after approximating
$V_3$ by evolving with only the tree diagrams and $\mybar{\mybar{V}}_3$. This
significantly reduces the errors in $E_\beta$
from simply dropping $V_3$ for both weak and
strong coupling. This is consistent with the observations above about the small
relative strength of the one-loop and two-loop contributions. Finally, in fig.
\ref{fig:1loop_v3_error} we show these errors when $V_3$ is evolved using both
tree and one-loop diagrams, dropping only the two-loop diagrams. This further
reduces the errors for weak coupling but not for strong coupling, where we have
seen in fig.~\ref{fig:v3ab_rhs} that the two-loop contribution, although smaller
than the one-loop contribution, becomes comparable to the contribution of
$\mybar{\mybar{V}}_3$ throughout the evolution.

It should be clear that this analysis is easily extended to the four-boson
sector, where $V_4$ must be added. For weak coupling, we should again find that
tree diagrams which are ${\cal O}(V_2 V_3)$ will dominate in the evolution of
$V_4$. Since there are no diagrams contributing to $V_4$ built only from two
powers of $V_2$, $V_4$ is fed entirely by induced interactions and might remain
small even for strong two-body interactions, but explicit calculations are
required to find what happens at strong coupling.


\section{Summary}
\label{sec:summary}

In this paper, we have used a simple two-level model system to
illustrate how a unitary SRG consistently evolves two-body and
three-body interactions.  We have developed a digrammatic treatment of
the SRG equations that applies generally and have shown how it can be
used with a variety of examples in our model. However, such a model
might seem too schematic to offer insight into realistic problems. In
fact, the steps to the physics problem that originally motivated this
work, that of describing atomic nuclei from microscopic interactions,
are not so large.

The low-energy nuclear many-body problem  has been one of the most
important problems in physics for over seventy years, but it has
stubbornly  resisted a complete solution. In recent years, however, it
has become clear that a Wilsonian renormalization group perspective
leads to new conceptual and practical ways to attack this problem. The
primary tool Wilson introduced to renormalization group formalism is a
renormalization transformation that lowers a well-defined hamiltonian
resolution, a tool that is ideally suited to finding universal
interactions. Feed any interaction constrained by low-energy data into
a well-designed RG transformation, it produces a
universal low-energy interaction that is decoupled from the ambiguous
(and irrelevant) high-energy components of the input interaction.

The same RG transformation  that Wilson employed in his first
non-perturbative RG calculation~\cite{wilson70} has recently been
applied to a wide variety of modern nucleon-nucleon
interactions~\cite{Vlowk2, VlowkRG} and every one of them was
transformed into a nearly universal low-energy effective
nucleon-nucleon interaction, called $\vlowk$.  It turns out that
$\vlowk$ also drastically reduces the complexity of the nuclear
many-body problem but requires the consistent evolution
of three-nucleon interactions~\cite{Vlowk3N}. 

Realistic nucleon-nucleon (NN) 
interactions include strong short-range repulsion and a strong tensor force. These lead to far
off-diagonal strength in the momentum representation of the nucleon-nucleon
interaction, so that large matrices are required to represent this interaction.
The size of these matrices increases rapidly with particle number.
$\vlowk$ requires
drastically fewer matrix elements to accurately reproduce all low-energy
NN scattering data, and this reduction exponentiates as we move to
the many-nucleon problem~\cite{NCSMSRG}. In contrast to
conventional NN interactions, which are highly nonperturbative, 
$\vlowk$ leads to nearly
converged nuclear matter calculations at second order in many-body
perturbation theory.  The three-nucleon interaction is an
essential ingredient
in saturating nuclear matter close to the empirical density and
binding energy~\cite{Bogner_nucmatt}. 
Even more, the
problem is not well-posed until consistent few-body interactions are found, 
as the
nucleon-nucleon interaction on its own does not maintain unitarity in even the
three-nucleon problem and violations are comparable to nuclear level spacings
\cite{NCSMSRG}.

The SRG transformation we
employ produces basically the same universal nucleon-nucleon
interaction as the $\vlowk$ evolution and shares the same
positive features.
Perhaps the most important advantage of the type of SRG transformation 
used here is that it ``automatically" allows us to consistently
evolve two- and few-nucleon interactions. Just as we have seen in our
model, if the
3-body force is not included in the evolution, few-body observables are cutoff
dependent and errors can be large. This article represents a first step
towards the computation of a fully consistent low-momentum 
three-nucleon interaction.
In addition to the
three-nucleon interaction that is required by unitarity when the two-nucleon
interaction flows, there is an intrinsic three-nucleon interaction that must be
added; current versions include various phenomenological and 
chiral effective field theory potentials. 
Ideally the SRG transformation
will again reveal a universal low-momentum, three-nucleon interaction. 

In principle, four-nucleon interactions can also be included
and may be needed for quantitative nuclear structure, particularly for
heavy nuclei.
The three-loop diagrams in the SRG equations
represent nine-dimensional integrals and the spin-isospin
algebra also grows in complexity. 
But we have seen in our model that even for strong
couplings we can sometimes accurately approximate the SRG evolution equations
by dropping such multi-loop contributions and we may even obtain better than
10\% accuracy using tree diagrams alone; 10\% accuracy for an already small
correction might  be sufficient. There are no integrals in these tree
diagrams, so we need ``only" deal with the spin-isospin algebra for four
nucleons. But this is idle speculation until the three-body interaction is
computed accurately. With a three-nucleon interaction in hand we can directly
test the important of $V_4$ by studying violations of unitarity in $^4$He, for
example.

The basic procedure for evolving two-nucleon and three-nucleon interactions
with a unitary RG transformation is the same as that illustrated in this
article. The square well must be removed, bosons must be replaced by nucleons,
one dimension goes to three dimensions and the initial hamiltonian must include
realistic inter-nucleon interactions. Removing the square well and replacing
bosons with fermions is uncomplicated, while moving up to three dimensions and
using realistic interactions requires more work but no major conceptual
developments.

It is important to realize that in nuclear calculations, we do not intend to
use the transformation to diagonalize the hamiltonian as we have in this
article. In realistic many-body problems we run the flow parameter $s$ only to
the point where low- and high-energy degrees of freedom are decoupled in the
few-nucleon sectors. Running the transformation beyond this point would produce
long-range few-body forces and strong many-body forces; we would effectively be
forced to {\emph solve} both few- and many-nucleon problems completely. We want
to simplify the few-body forces that are then employed in many-nucleon
calculations, not diagonalize the hamiltonian in every sector of Fock space and
produce the large tower of many-body forces that would be required to do this
in the many-body sectors. When we achieve sufficient decoupling in the
two-nucleon sector, we compute the three-nucleon interaction using methods
illustrated in this article and then switch to conventional  few- or many-body
methods using this partially diagonalized hamiltonian. See
Refs.~\cite{Bogner:2006srg} and \cite{Bogner:2007jb} for details.

Perhaps the most interesting immediate extension of the model we present is to
remove the square well and let $T$ simply be the kinetic energy. Instead of
discrete sums, we obtain integrals over continuous momenta, but these can be
approximated using a discrete mesh, so we once again have sums over discrete
indices. $N$ bosons interacting via a two-body
delta-function potential in one dimension produce one bound state whose wave
function and energy are known analytically \cite{exact_1d_boson}. Using
momentum eigenstates we must compute the evolution of $\{k_1,k_2 \mid V_2 \mid
k_1',k_2'\}$ and $\{k_1, k_2, k_3 \mid V_3 \mid k_1', k_2', k_3'\}$. These can
then be used in a numerical basis function calculation that should converge
increasingly rapidly with basis size to the analytic result as the unitary
transformation is run. With this discrete basis, 
the evolution equations are almost identical to those in this article.

The basic unitary renormalization group equation we employ in this
article,  eq.~(\ref{eq:commutator}), is the same in three dimensions as
in one. 
Replacing bosons with fermions means only minor changes in the
diagrammatic rules for the evolution equations, with the familiar minus signs
from fermion exchange and no symmetry factors associated with loops.
Exploiting angular momentum algebra and choosing appropriate
Jacobi coordinates for three-nucleon states introduce technical
complications, but these appear in a form almost identical to what is
faced when solving three-nucleon Faddeev equations \cite{fewbody}. 
The SRG equations governing the evolution of the nuclear three-body force
are only a scaled-up version (larger matrices) of our model problem,
as opposed to the multi-channel complications of
the Faddeev equations.

In this article we have chosen one simple operator for $T$, but there are many
possible choices and this freedom can be exploited to improve convergence in
many-body calculations. The calculations in this article are easily modified
for any $T$ that can be defined by diagonal matrix elements in the square well
basis. $T$ can itself depend on the flow parameter. Wegner \cite{Wegner:1994}
advocated using the diagonal matrix elements of $H_\flow$ itself, which
corresponds in the $2\times 2$ matrix model  to choosing $T=\omega_z(s) \sigma_z$.
Wegner's transformation is guaranteed to drive the momentum representation of
the hamiltonian to diagonal form, while other choices can produce divergent
matrix elements and stall.
These considerations are discussed in the context of simple models in
Ref.~\cite{Glazek:2007}.


\begin{acknowledgments}
We thank  E.~Anderson, S.~Glazek, E.~Jurgenson, and A. Schwenk
for useful comments. 
This work was supported in part by the National Science 
Foundation under Grant Nos.~PHY--0354916, PHY-0456903, and PHY--0653312 
and by the UNEDF SciDAC Collaboration under DOE Grant DE-FC02-07ER41457.
\end{acknowledgments}

\end{document}